\documentclass[letter,10pt]{article}
\pdfoutput=1
\usepackage{jheppub}

\usepackage[usenames,dvipsnames]{xcolor}

\usepackage{epsf,epsfig,subfigure,graphicx,amsmath,amssymb,mathtools,hhline}
\usepackage{color}
\usepackage{float}
\usepackage{multirow}
\usepackage{nicefrac}
\usepackage{epstopdf,slashed}
\usepackage[normalem]{ulem}

\makeatletter

\newcommand{\fmslash}[2][0mu]{%
  \mathchoice
    {\fmsl@sh\displaystyle{#1}{#2}}%
    {\fmsl@sh\textstyle{#1}{#2}}%
    {\fmsl@sh\scriptstyle{#1}{#2}}%
    {\fmsl@sh\scriptscriptstyle{#1}{#2}}}
\newcommand{\fmsl@sh}[3]{%
  \m@th\ooalign{$\hfil#1\mkern#2/\hfil$\crcr$#1#3$}}
\makeatother

\newcommand{\lsim}{{\;\raise0.3ex\hbox{$<$\kern-0.75em\raise-1.1ex\hbox{$\sim$}}\;}}
\newcommand{\gsim}{{\;\raise0.3ex\hbox{$>$\kern-0.75em\raise-1.1ex\hbox{$\sim$}}\;}}

\newcommand{\beq}{\begin{equation}}
\newcommand{\eeq}{\end{equation}}
\newcommand{\bea}{\begin{eqnarray}}
\newcommand{\eea}{\end{eqnarray}}
\mathchardef\minus="002D

\usepackage{color}

\def\beq{\begin{equation}}
\def\eeq{\end{equation}}
\def\bea{\begin{eqnarray}}
\def\eea{\end{eqnarray}}

\makeatletter
\newcommand\footnoteref[1]{\protected@xdef\@thefnmark{\ref{#1}}\@footnotemark}
\makeatother

\title{Boosted Dark Matter Quarrying at Surface Neutrino Detectors}

\author[a]{Doojin Kim,}
\author[b]{Kyoungchul Kong,}
\author[c]{Jong-Chul Park,\footnote{\label{note1}Corresponding authors.}}
\author[d,e]{Seodong Shin\footnoteref{note1}}

\affiliation[a]{Theoretical Physics Department, CERN, Geneva, Switzerland}
\affiliation[b]{Department of Physics and Astronomy, University of Kansas, Lawrence, KS 66045, USA}
\affiliation[c]{Department of Physics, Chungnam National University, Daejeon 34134, Republic of Korea}
\affiliation[d]{Enrico Fermi Institute, University of Chicago, Chicago, IL 60637, USA}
\affiliation[e]{Department of Physics \& IPAP, Yonsei University, Seoul 03722, Republic of Korea}
\emailAdd{doojin.kim@cern.ch, kckong@ku.edu, jcpark@cnu.ac.kr, seodongshin@yonsei.ac.kr}

\preprint{
\begin{minipage}{3.5cm}
\small
CERN-TH-2018-088 \\
EFI-18-5\\
\end{minipage}
}

\abstract{
We propose the idea of ``Earth Shielding'' to reject cosmic-ray backgrounds, in the search for boosted dark matter at surface neutrino detectors, resulting in the enhancement of the signal-to-background ratio.
The identification of cosmic-originating rare signals, especially lacking features, at surface detectors is often considered hopeless due to a vast amount of cosmic-ray-induced background, hence underground experiments are better motivated to avoid such a challenge.
We claim that surface detectors can attain remarkable sensitivities to even featureless signals, once restricting to events coming through the Earth from the opposite side of the detector location for the signals leaving appreciable tracks from which the source direction is inferred.
By doing so, potential backgrounds in the signal region of interest can be substantially suppressed.
To validate our claim, we study experimental reaches at several surface experiments such as SBN Program (MicroBooNE, ICARUS, and SBND) and ProtoDUNE for elastic boosted dark matter signatures stemming from the Galactic Center.
We provide a systematic discussion on maximizing associated signal sensitivities.
}

\begin{document}

\maketitle

\section{Introduction}

The existence of dark matter (DM) in the universe and non-vanishing masses of neutrinos are most robust, empirical pieces of evidence advocating the presence of new physics beyond the Standard Model (SM).
Therefore, the exploration into the associated physics sectors offers an excellent avenue of deepening our understanding in particles physics.
The Large Hadron Collider, an energy-frontier experiment, has been playing the role of a major steering gear in the search for relevant signals.
For example, the observation of large missing energy in association of a single visible particle is one of the well-motivated DM search strategies.
However, we remark that the signal rates associated with neutrinos or DM are typically small, due to their elusive nature.
Therefore, experimental approaches with highly intensified particle beams such as fixed target experiments are often motivated in order to increase signal statistics, as far as the mass scale of relevant particle of interest is within the reach of accelerator beam energy.
An alternative experiment scheme is to wait for signals coming from various astrophysical sources, e.g., the Galactic Center (GC), the Sun, dwarf spheroidal galaxies (dSphs), and study relevant phenomena (henceforth called ``cosmic-frontier'' approach), placing large-volume experimental apparatuses in a desirable location.

Relevant detectors are typically installed deep under the ground mainly in order to avoid an overwhelming amount of cosmic-ray-induced background or noise to the signals of interest.
In this sense, it is common lore that fulfilling the same sort of experiments with detectors near the surface is somewhat nonsensical because signal candidate events would get buried inside the cosmic backgrounds~\cite{Goodenough:2012mj}.
Of course, if a signal process accompanies many unique features, it is possible to reject cosmic background events so sufficiently that the above-mentioned challenge may be mitigated~\cite{Chatterjee:2018mej}.
By contrast, if it is nearly featureless, a certain fraction of cosmic-ray events can fake the signal very easily because of, for example, particle misidentification, hence it may be hopeless to perform interesting phenomenological studies with the detectors located either on the ground or close to the ground (which we collectively call surface or surface-based detectors throughout this paper).
Nevertheless, if a given detector enables to infer the direction of signal sources, one may be allowed to focus on the events incoming along a specific direction.
Furthermore, if potential backgrounds in the direction are significantly suppressed while signals are intact, the associated experiments may acquire enough signal sensitivity.

Along this line, we point out that the Earth itself can play a role of shielding the cosmic rays coming through from the opposite side, with respect to the surface detectors.
The notion of utilizing the Earth itself as a cosmic-ray blocker is well known for a long time, having encouraged to place detectors deep underground as mentioned earlier.
The idea of restricting to events coming along particular directions on top of it has sometimes been employed to enhance signal purity; for example, various underground, underwater, and under-ice experiments such as Super-Kamiokande (SK)~\cite{Choi:2015ara}, AMANDA~\cite{Ackermann:2005fr}, ANTARES~\cite{Adrian-Martinez:2013ayv}, and IceCube~\cite{Aartsen:2016zhm} observed the upward-going lepton signals as a way of DM indirect detection.
As we will demonstrate throughout this paper, however, the application of the idea to surface-based detectors gives rise to quite striking outputs in the sense of opening unexpected physics opportunities.
Furthermore, given that an increasing number of (relatively) large-volume surface-based detectors are in operation or planned these days, we believe that thinking of a similar experiment strategy, dubbed ``Earth Shielding'', for them is rather timely and highly motivated.
Indeed, among the surface-based neutrino experiments, the NuMI Off-axis $\nu_e$ Appearance (NO$\nu$A) Collaboration has attempted to apply this idea for the observation of upward-going muons induced by muon neutrinos created by DM annihilation at the Sun~\cite{Mina:2015efa}, although official physics results are not available yet.
The major background to their signal is the upward-going muon flux, created underground, which can be efficiently vetoed by its track information recorded by the detector in combination with their trigger system.

In this paper, we generalize the idea to any generic signal events which can leave sizable track(s) at a surface detector so that only those coming through the Earth are accepted as valid events.
As a result, any reducible, signal-faking, cosmic-ray-induced backgrounds are significantly suppressed, and irreducible backgrounds such as neutrino-initiated events remain as the dominant background like in usual underground experiments.
Practically, we propose to use the data coming out of the plane extended by the bottom of the surface detectors for a certain time window, which may differ day-by-day, depending on the interrelation among the signal-source point, the Earth rotation axis, and the detector location.
For example, if a surface detector located in the south pole searches for the signal coming from the Sun,  24-hr (0-hr) data out of the bottom plane is usable in summer (winter) of the northern hemisphere.
This method is also inspired from the fact that the signal-to-background ratio {\it modulates daily} because of the rotation of the Earth and becomes maximized when the signal origin is facing the Earth surface opposite to the detector location in principle.

The idea of Earth Shielding is very general so that it can be used for various signals. 
As a concrete example to give rise to featureless signal events, the (minimal) two-component boosted dark matter (BDM) scenario~\cite{Agashe:2014yua} is taken into account throughout the rest of this paper. 
In typical cases, the heavier component dominates over the cosmological dark matter~\cite{Belanger:2011ww}, and can pair-annihilate into the lighter in the universe today.
The mass gap between the two DM components allows a significant boost factor for the lighter, so that it may leave a relativistic scattering signature at detectors.
More specifically, if it scatters off electron in detector material, the existence of such an event can be inferred {\it only} from the resulting electron recoil which is easily mimicked by cosmic rays as explained earlier.

We organize the flow of this paper as follows. In the first half of Sec.~\ref{sec:surface}, some existing and projected candidate surface detectors are listed along with their key characteristics.
The last half is devoted to discussing potential cosmic-ray backgrounds with careful estimations on them.
In Sec.~\ref{Earth Shielding}, we elaborate our proposal of ``Earth Shielding'', followed by the discussion on a way of improving the signal-to-background ratio.
We then list up a few representative physics scenarios, to which the proposed signal search strategy is applicable, in Sec.~\ref{sec:scenarios}. Of them, we select a benchmark scenario, for which our phenomenological studies are performed, and give a brief, theoretical review on it.
Sec.~\ref{sec:analysis} contains several experimental reaches set by the events collected according to our proposal here.
Finally, we reserve Sec.~\ref{sec:conclusions} for conclusions and outlook.


\section{Surface Experiments and Cosmic Rays \label{sec:surface}}

Several existing or projected experiments have detectors installed (almost) on the ground.
Most of them are aiming at neutrino physics as their first priority mission, in conjunction with high-intensity proton beam sources.
Examples include NO$\nu$A~\cite{Ayres:2007tu}, Micro Booster Neutrino Experiment (MicroBooNE)~\cite{Antonello:2015lea, Acciarri:2016smi}, Imaging Cosmic And Rare Underground Signals (ICARUS)~\cite{Antonello:2015lea},\footnote{The ICARUS T600 detector~\cite{Amerio:2004ze} started first operation in 2010 at Gran Sasso, and moved to Fermi National Laboratory as part of the Fermilab Short-Baseline Neutrino (SBN) Program~\cite{Antonello:2015lea}.} Short-Baseline Near Detector (SBND)~\cite{Antonello:2015lea}, and a prototype of far detectors at Deep Underground Neutrino Experiment (ProtoDUNE)~\cite{Abi:2017aow, Agostino:2014qoa}.
The first two are currently collecting data, while the last three are expected to be in operation within a few years.
All of the above detectors have been usually considered {\it not} ideal for signal searches at the cosmic frontier again due to the challenge from a tremendous amount of cosmic-ray background.
In particular, ProtoDUNE is not originally intended to pursue physics opportunities partly due to the cosmic background issue, but puts its high priority mission on validating and testing the technologies and designs that are adopted by the DUNE far detectors.\footnote{Note, however, that
Ref.~\cite{Chatterjee:2018mej} has recently pointed out the potential of physics opportunities at ProtoDUNE in the context of DM physics, for the first time.}

We shall show that the above-mentioned experiments may have emergent cosmic-frontier physics potentials with the aid of the Earth shielding to resolve the cosmic background issue.
Therefore, it is instructive to highlight some of the detector characteristics of those experiments for later use.
\begin{itemize}
\item NO$\nu$A~\cite{Ayres:2007tu}: The detector is placed almost on the ground up to overburden of barite and concrete, containing liquid scintillators (LS) in a cellular structure, and has been in full operation since October 2014 with 14 kt active mass.
    It (relatively) lacks official analyses on angular resolution ($\theta_{\rm res}$) and threshold energy ($E_{\rm th}$) for upward-going electrons (from the Earth direction).
\item MicroBooNE~\cite{Antonello:2015lea, Acciarri:2016smi}, ICARUS~\cite{Antonello:2015lea}, SBND~\cite{Antonello:2015lea}: As individual experiments in the SBN Program, MicroBooNE has been in operation since July 2015 while the other two will run within a few years. All detectors employ the Liquid Argon Time Projection Chamber (LArTPC) technology. They are/will be placed about 6 m underground. Their total liquid-phase Argon masses are 170 t, 760 t, and 220 t while active masses are 89 t, 476 t, and 112 t, respectively.
\item ProtoDUNE~\cite{Abi:2017aow, Agostino:2014qoa}: Two types of LArTPC detectors, Single-phase (SP) and Dual-phase (DP), are under construction at CERN. They consist of about 1.5 kt of total liquid-phase Argon mass (770 t for SP and 705 t for DP), and about 720 t of active mass (420 t for SP and 300 t for DP).
\end{itemize}
We also tabulate key values in Table~\ref{tab:sumdetector} for convenience of reference.
In this work, however, we will focus only on LArTPC detectors, as the information for electrons incident in the Earth center direction is relatively better known or inferred.
We leave the study with NO$\nu$A for future.

\begin{table*}
\centering
\scalebox{0.765}{
\begin{tabular}{@{} c | c  c  c  c  c  c  c @{}}
\hline \hline \\ [-1em]
\multirow{2}{*}{Detector} & Target & \multicolumn{2}{c}{Active volume} & Fiducial volume & \multirow{2}{*}{Depth} & \multicolumn{2}{c}{Electron} \\
 & material & $w \times h \times l$ [m$^3$] & mass [kt] & mass [kt] &  & $E_{\rm th}$ [MeV] & $\theta_{\rm res}$ \\ [0.1em]
\hline \\ [-1em]
\multirow{2}{*}{NO$\nu$A} & PVC cells & \multirow{2}{*}{$15.5 \times 15.5 \times 60$} & \multirow{2}{*}{14} & \multirow{2}{*}{--} & 3 m overburden of & \multirow{2}{*}{unclear} & \multirow{2}{*}{unclear} \\
 & filled with LS &  &  &  & concrete \& barite &  &  \\  [0.1em]
MicroBooNE & LArTPC & $2.56 \times 2.33 \times 10.37$ & 0.089 & 0.055 & $\sim$ 6 m underground & $\mathcal{O}(10)$ & $\mathcal{O}(1^\circ)$ \\  [0.1em]
ICARUS & LArTPC & $2.96 \times 3.2 \times 18$ ($\times 2$) & 0.476 & $\sim0.3$ & $\sim$ 6 m underground & $\mathcal{O}(10)$ & $\mathcal{O}(1^\circ)$ \\  [0.1em]
SBND & LArTPC & $4 \times 4 \times 5$ & 0.112 & $\sim0.07$ & $\sim$ 6 m underground & $\mathcal{O}(10)$ & $\mathcal{O}(1^\circ)$ \\  [0.1em]
ProtoDUNE SP & LArTPC & $3.6 \times 6 \times 7$ ($\times 2$) & $\sim0.42$ & $\sim0.3$ & on the ground & $\sim30$ & $\sim1^\circ$ \\  [0.1em]
ProtoDUNE DP & LArTPC & $6 \times 6 \times 6$ & $\sim0.3$ & $\sim0.21$ & on the ground & $\sim30$ & $\sim1^\circ$ \\  [0.1em]
\hline \hline
\end{tabular}
}
\caption{\label{tab:sumdetector}
Summary of key characteristics of several surface detectors.
See the text for the detailed explanation.
The numbers with the ``$\sim$'' symbol are our estimations based on those of similar detectors due to the lack of official announcement.
``$\times 2$'' in parentheses indicates that the relevant detector is composed of two consecutive equal-sized sections.
}
\end{table*}

In the analysis of Ref.~\cite{Abratenko:2017nki}, the borders of the fiducial volume of MicroBooNE are defined as 20 -- 37 cm inward from each border of the active volume, and the resulting fiducial mass is reported as 55 t.
For ProtoDUNE~\cite{Chatterjee:2018mej}, we set back 35 cm from the borders of the active volumes following DUNE conceptual design report Vol.~IV~\cite{Acciarri:2016ooe}, which results in $\sim$300 t and $\sim$210 t of masses for the fiducial volumes of the SP and DP, respectively.
Considering the situation for MicroBooNE and DUNE/ProtoDUNE, we believe that it is fairly reasonable to assume that the fiducial volumes of ICARUS and SBND are defined as 30 cm inward from the borders of the active volumes, which leads to the fiducial masses of $\sim$300 t and $\sim$70 t, correspondingly.

For the three experiments in the SBN Program (MicroBooNE, ICARUS, and SBND),
no official detection threshold energy and angular resolution values for the electron signal currently exist.
Only indirect information on $E_{\rm th}$ is available for MicroBooNE~\cite{Woodruff:2017kex, Acciarri:2017sjy}, while the experiments adapting a similar LArTPC technology such as ArgoNeuT~\cite{Acciarri:2016sli} and DUNE/ProtoDUNE~\cite{Abi:2017aow, Acciarri:2015uup, Agostino:2014qoa} have provided their expected $E_{\rm th}$ and $\theta_{\rm res}$.
We conduct our analysis, assuming that the unknown $E_{\rm th}$ and $\theta_{\rm res}$ of MicroBooNE, ICARUS, and SBND are the same as the expected ones of DUNE/ProtoDUNE~\cite{Private}, i.e., $E_{\rm th}=30$ MeV and $\theta_{\rm res}=1^\circ$ unless otherwise noted.\footnote{In fact, $\theta_{\rm res} \approx 10^\circ$ is enough for the all the applicable scenarios suggested in Sec.~\ref{sec:pheno}, except point-like sources (e.g. boosted DM from the Sun).
In the example data analysis, we consider a scenario in which the source is not point-like.
Thus, even with much worse $\theta_{\rm res}$, e.g., $\sim10^\circ$, our results presented in Sec.~\ref{sec:pheno} are robust.}

Moving onto the cosmic-ray background, we divide our discussion into neutrinos and non-neutrino particles.
The latter is further divided into muon and other high energy particles including electron/positron and pion.
\begin{itemize}
\item {\bf Atmospheric neutrinos.} Neutrinos mostly leave their scattering signatures with matter {\it inside} the ``fiducial'' volume of a detector, not generating tracks coming from the outside.
This feature resembles what is expected from the DM interaction, hence they are often considered as irreducible background in DM searches.
The experiments of our interest are designed with threshold energies greater than $\mathcal{O}(10\hbox{ MeV})$ so that they do not have sensitivity to solar neutrino events but only to the atmospheric neutrinos.
The flux of the latter is substantially smaller than that of the former, and comes into the detector almost {\it isotropically}.
We now discuss the number of expected atmospheric electron neutrino-induced events $N_{\nu_e{\rm -ind}}^{\textrm{atm}}$, based on the study by the DUNE Collaboration~\cite{Acciarri:2015uup}.
From the fully contained electron-like sample, we estimate that for recoil electron energy $E_e \geq 30$  MeV\footnote{As a matter of fact, Ref.~\cite{Acciarri:2015uup} does not clearly state the range of $E_e$ that they considered. So, we here assume that any $E_e$ greater than $E_{\rm th}(\sim 30 \hbox{ MeV})$ was considered in the study. }
\bea
N_{\nu_e{\rm -ind}}^{\textrm{atm}} \approx 40.2\, \textrm{yr}^{-1}\textrm{kt}^{-1}\,, \label{eq:nuflux}
\eea
which is in good agreement with the estimation based on the fully contained $e$-like, single-ring, 0-decay electron, and 0-tagged neutron events reported by the SK Collaboration~\cite{Abe:2014gda,Kachulis:2017nci}.
Note that for the above simulation and measurement the effect from neutrino oscillations was taken into account.

\item {\bf Muon.} On the other hand, non-neutrino particles including electrons, muons, and pions are, in principle, reducible.
The real problem comes about due to their humongous fluxes; even if a tiny fraction of events are mistakenly identified/accepted as signal events, the resulting number of background events will be huge.
For example, the integral density of vertical muons with energy greater than 1 GeV at sea level is $\sim 70\, \textrm{m}^{-2}\textrm{s}^{-1}\textrm{sr}^{-1}$~\cite{Patrignani:2016xqp}. Considering the observation that the muon energy spectrum appears almost flat below 1 GeV~\cite{Patrignani:2016xqp}, we estimate the integrated vertical muon flux $N_{\mu}$ above $\sim 10$ MeV to be
\bea
N_\mu \approx 100\, \textrm{m}^{-2}\textrm{s}^{-1}\textrm{sr}^{-1}\approx 10^{10}\, \textrm{m}^{-2}\textrm{yr}^{-1} \label{eq:muonflux}
\eea
with $\pi$ steradians included.\footnote{Strictly speaking, the muon flux depends on the latitude, but such a dependence is not large enough to alter the argument hereafter.}
More recently, the MicroBooNE Collaboration has estimated that a few hundred muons per square meter per second, which is comparable to the estimate in~\eqref{eq:muonflux}, may cross the detector active volume~\cite{Acciarri:2017rnj}.
Cosmic muon taggers may be supplemented to a detector for a more efficient muon veto, but an impractically small missing rate is required to suppress cosmic muons to a manageable level.

\item {\bf Other high-energy particles.} The fluxes of other high-energy cosmic particles such as electron/positron and pion are sub-leading as they are smaller by about $3-4$ orders of magnitude than the muon flux~\cite{Patrignani:2016xqp}.
However, their stopping powers in material are much larger than that of the muon. Therefore, a conservative definition of the detector fiducial volume can significantly abate their contributions.
Finally, we discuss cosmic neutrons.
The neutron flux is smaller by about two orders of magnitude than the muon one, so the resultant contribution is not negligible.
The coupling of GeV-range neutrons to matter is mostly mediated by the strong force, thus they rapidly break up in material.
By contrast, MeV-range neutrons can reach the detector fiducial volume without leaving any trace and (predominantly) scatter off nuclei.
Therefore, if the signal of interest involves only electron recoil, one may avoid the neutron-induced backgrounds.
\end{itemize}
In summary, among the above-listed backgrounds, the muon-induced is most challenging in surface-based experiments due to its formidable flux and less handles to reject it, although reducible.
In this sense, the idea of Earth shielding, which we shall elaborate in the next section, is mainly targeting at the muonic background. 
We shall show that indeed the (irreducible) neutrino-induced background remains dominant as in the case of the underground experiments.


\section{Earth Shielding}\label{Earth Shielding}

In this section, we describe our proposal of ``Earth Shielding'' in surface-based experiments to observe a DM-induced scattering signal off target even when its kinematic topology is featureless, starting by explaining the main idea.
A brief discussion on how to improve signal sensitivity via angular cuts follows in the next subsection.
Benefiting from the ``Earth Shielding'' depends on the relation among the signal-source direction, Earth's rotation axis, and the detector latitude, so we explicate a way of calculating effective exposure in the last subsection, taking a couple of examples that signals come from the Sun and the GC.

\subsection{Main idea}
The key idea is to use the Earth as shielding material against cosmic-ray backgrounds; one takes the events coming only from the bottom of the surface based detectors for a certain period of time, depending on the region where the signal of interest is originating from.
The basic concept of observing the upward-going signals has been widely used in the DM indirect detection via its annihilation into the SM neutrino pair coming from the GC or Sun, in various underground, underwater, or under-ice detectors such as SK~\cite{Choi:2015ara}, AMANDA~\cite{Ackermann:2005fr}, ANTARES~\cite{Adrian-Martinez:2013ayv}, and IceCube~\cite{Aartsen:2016zhm}.
When it comes to the surface-based neutrino experiments, the NO$\nu$A Collaboration embarked on a project to observe upward-moving muons~\cite{Mina:2015efa}, which is conceptually similar to our proposal, although follow-up experimental results (including the issues related with the upward-moving muon trigger) are still unannounced.

When muons penetrate the Earth, the intensity of vertical muons ($N_\mu \approx 100\, \textrm{m}^{-2}\textrm{s}^{-1}\textrm{sr}^{-1}$ at sea level) decreases rapidly with depth and becomes flat ($N_\mu \approx 2\times10^{-9}\, \textrm{m}^{-2}\textrm{s}^{-1}\textrm{sr}^{-1}$) at depth around $\sim$20 km.w.e..\footnote{We quote the depth as kilometer water equivalent (km.w.e.), a standard measure of cosmic ray attenuation, which is defined as the product of depth (in km) and density of material (here the Earth crust) relative to that of water.}
This flatness is due to muons produced locally by charged-interactions of $\nu_\mu$ \cite{Patrignani:2016xqp}.
If a muon approaches a detector from underneath at an angle $\phi$ with respect to horizontal surface where a detector is located, the muon loses energy while getting through the crust of the Earth by distance $2 R_\oplus \sin\phi$. Here $R_\oplus$ denotes the radius of the Earth whose value is 6371 km.
If the traveling distance is longer than $d_{\rm flat}\approx 20 \, {\rm km.w.e.} \approx\,$7 km at the Earth's crust, the vertical intensity would drop by a factor of $\sim10^{11}$.
In other words, $\phi$ should be greater than the corresponding critical value $\phi_{\rm cr}$ given as follows:
\bea
\phi_{\rm cr} = \sin^{-1}\frac{d_{\rm flat}}{2 R_\oplus} \approx \frac{d_{\rm flat}}{2 R_\oplus} \approx 0.03^\circ. \label{eq:cr}
\eea
Therefore, the number of upward-going muons with $\phi>\phi_{\rm cr}$ will be
\bea
N_\mu^{{\rm upward}} \approx 0.1~\textrm{m}^{-2} \textrm{yr}^{-1}. \label{eq:upwardmuon}
\eea
Moreover, such muons should enter the fiducial volume without leaving a track to mimic signals.
We can conservatively estimate the probability of such ``sneaking-in'' muons to be $\sim 10^{-3}$~\cite{Chatterjee:2018mej, Acciarri:2017rnj,MBEpub}.
Combining this probability and the estimate in~\eqref{eq:upwardmuon}, we can safely ignore upward-going muon backgrounds. This further means that the dominant backgrounds would be atmospheric neutrinos entering a detector isotropically.
Since we now take the hemisphere underneath a detector at surface level, the corresponding numbers of both signal and neutrino events should be reduced by a factor of 2.

\subsection{Sensitivity improvement by an angular cut}
\begin{figure*}[t!]
\centering
\includegraphics[width=0.47\linewidth]{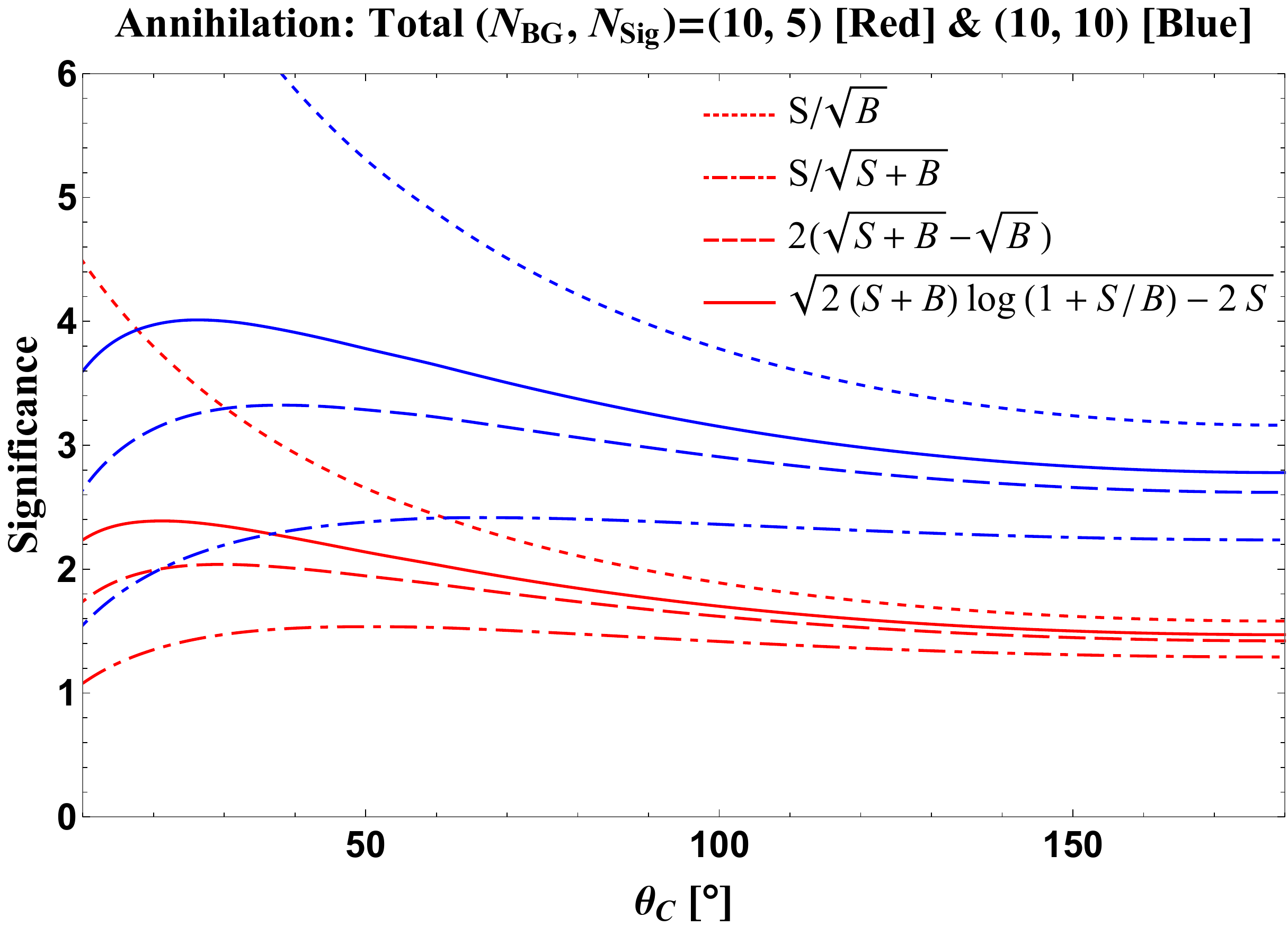}
\hspace{0.1cm}
\includegraphics[width=0.48\linewidth]{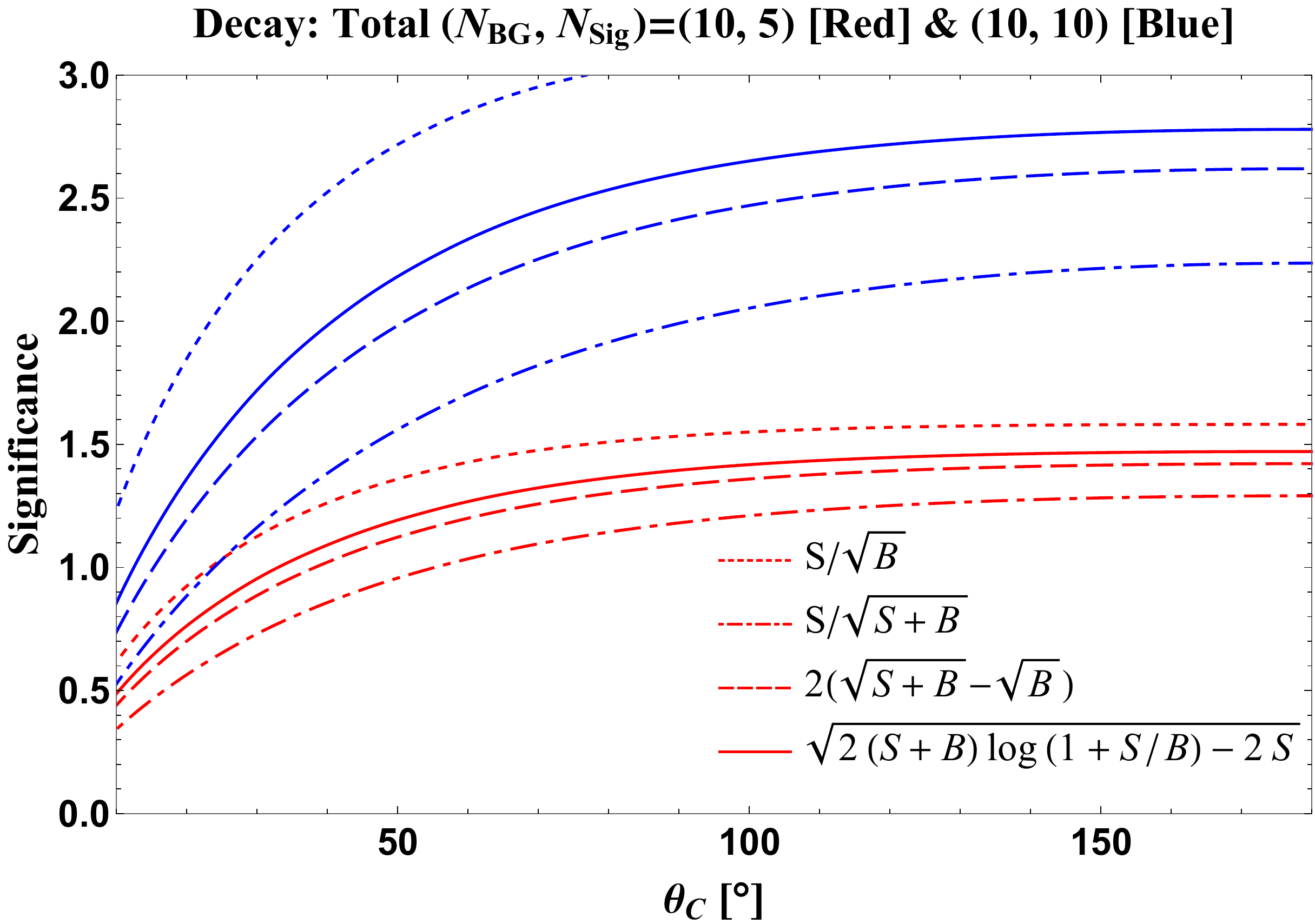}
\caption{\label{fig:Sig-Compare}
Signal significance as a function of search cone angle $\theta_C$ for various calculation methods.
Signals originate from DM annihilation (left panel) and DM decay (right panel).
The total expected numbers of background and signal events from all sky are normalized to $(N_{\rm BG}, N_{\rm Sig}) =$ (10, 5) and (10, 10) for red and blue curves, respectively.
}
\end{figure*}

LArTPC-based detectors have great advantage over others, e.g., Cherenkov-based detectors, in terms of not only lower threshold and better angular resolution but excellent particle identification, hence providing an additional reduction in atmospheric neutrino backgrounds.
With the assumption of a uniform flux, the number of atmospheric neutrino background events within a cone of angle $\theta_C$ around the direction of the DM source obeys the following relation:
\bea\label{eq:Nallsky}
N_{\rm BG}(\theta_C) = D_f \sin^2\frac{\theta_C}{2} N_{\rm BG}(180^\circ)\,,
\eea
where the total number over all sky $N_{\rm BG}(180^\circ)$ is given by~\eqref{eq:nuflux}.
Be aware that we multiplied prefactor $D_f$ to reflect that we effectively consider a certain fraction of day when the Earth shielding effect can be utilized, with respect to the source core.
We shall discuss $D_f$ in great detail in the next subsection, but we set $D_f$ to be 1/2 (relevant to the case where the Sun is a signal source) throughout the rest of this subsection for illustration.

We can estimate the signal significance as a function of search cone angle $\theta_C$, with the major background rate given according to Eq.~\eqref{eq:Nallsky}.
To see the search cone angle dependence of the signal rate, we consider two typical scenarios: signals originating from DM annihilation and decay.
In our analysis, we choose the search cone centered on the direction to the GC assuming the Navarro-Frenk-White (NFW) halo profile~\cite{Navarro:1995iw, Navarro:1996gj}.\footnote{Other DM halo profiles can affect detailed values, but exactly the same analysis method should get through.}
In Ref.~\cite{Agashe:2014yua}, $\theta_C\simeq10^\circ$ was suggested as an optimal cone angle in the search for the boosted DM, which is created in the universe today via annihilation of dominant relic DM,\footnote{See also Sec.~\ref{sec:scenarios} for the underlying theoretical argument.} at large-volume ($\gg1$ kt) Cherenkov detectors.
This value essentially maximizes the significance defined as $N_{\rm Sig}(\theta_C)/\sqrt{N_{\rm BG}(\theta_C)}$, where $N_{\rm BG}(\theta_C)$ and $N_{\rm Sig}(\theta_C)$ are the numbers of background (or $B$) and signal (or $S$) events within a cone of angle $\theta_C$.
However, this formula of the expected signal significance is valid only when $N_{\rm BG} \gg N_{\rm Sig} \gg 1$, which is not the case for the LArTPC detectors with volumes $< 1$ kt under consideration.
Hence, we use various approximated formulas of expected signal significance in Ref.~\cite{Patrignani:2016xqp, Bityukov:1999sa} and show how they change in terms of $\theta_C$ to choose the optimal angle maximizing the significance for the BDM search.
As displayed in the left panel of Fig.~\ref{fig:Sig-Compare}, large search cone angle
$\theta_C$ around $20^\circ - 30^\circ$ is optimal for annihilating DM scenarios with a relatively low background rate.
Moreover, comparison of different signal events for a fixed number of background events $N_{\rm BG}$ (left panel) shows that the optimal search cone angle $\theta_C$ is larger if $N_{\rm Sig}$ (or equivalently, the signal-to-background ratio) is bigger.
We also explicitly check that the optimal angle $\theta_C$ is almost intact for a fixed model point just with more data accumulation through a larger exposure time and/or a larger detector volume.
On the contrary, the right panel clearly suggests that the optimal search angle $\theta_C$ is simply $180^\circ$, i.e., the whole sky, for a decaying DM scenario.
This is because the signal flux under the decaying DM is {\it linearly} proportional to the DM number density so that the increase of $N_{\rm Sig}$ with $\theta_C$ around the GC is not as rapid as that in the case of the annihilating DM with the flux quadratically proportional the DM number density.

\subsection{Effect of Earth's rotation}

\begin{figure}[t]
\centering
\includegraphics[width=15.5cm]{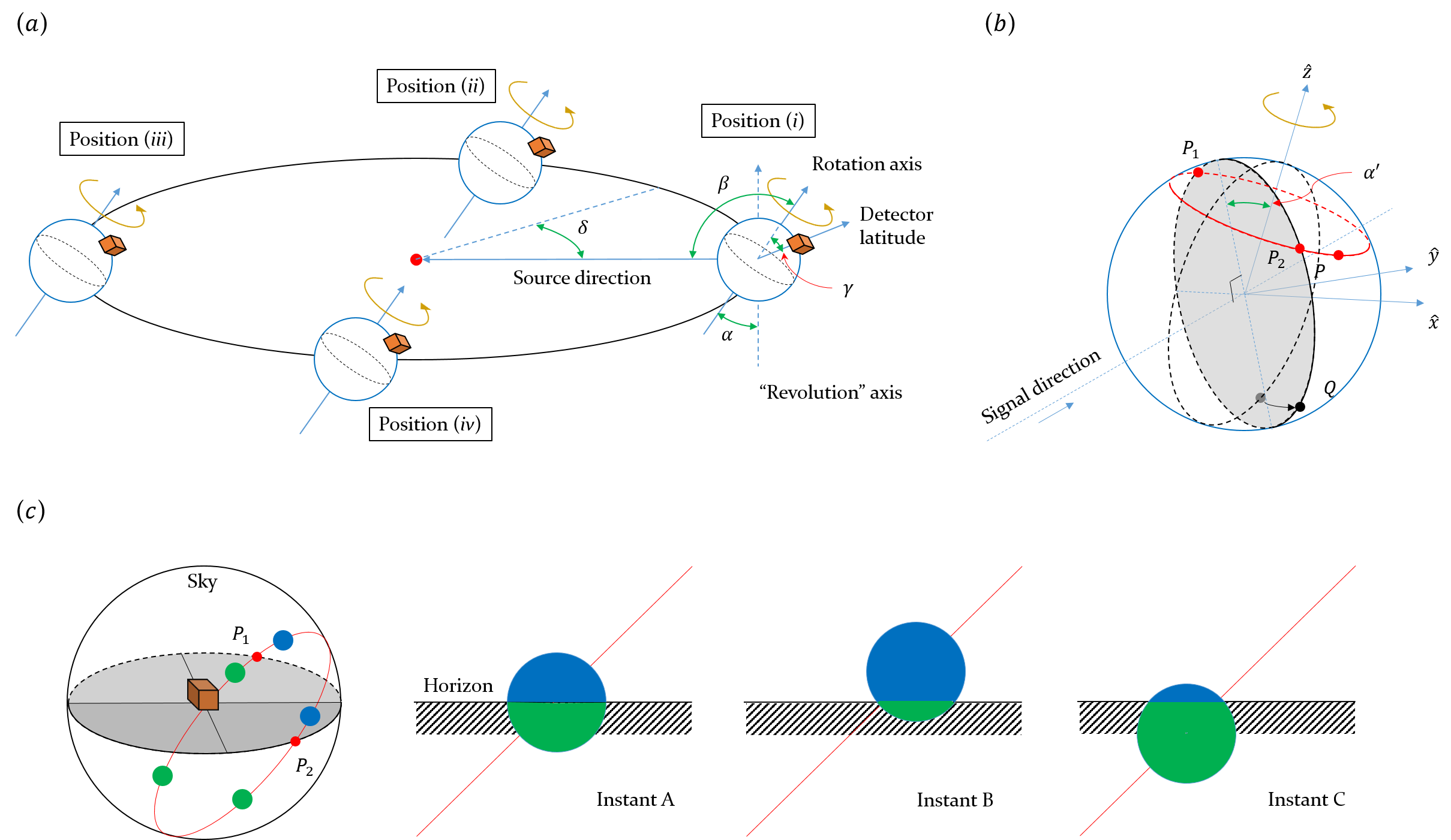}
\caption{\label{fig:rotation} Panel ($a$): Various positions of the Earth relative to a given signal source point.
Panel ($b$): Some important coordinates related to the ``Earth Shielding'' method. The given detector here can use data from $P_2$ through $P_1$.
Panel ($c$): The travel of a non-point-like, $\theta_C$-cone-spanned source on the celestial sphere according to the Earth's rotation. At instant A, the source core passes through the horizon, so a half of the cone (green area) benefits from the ``Earth Shielding''. 
On the other hand, at instant B (instant C), the source core is above (below) the horizon, so the smaller green (blue) cone benefits from (has no benefit from) the ``Earth Shielding''.
}
\end{figure}

As mentioned earlier, the amount of time during which signal events come through the Earth is, in general, governed by the source direction, the rotation axis, and the detector location.
To get a more quantitative understanding, let us imagine that the Earth revolves around a given source point as shown in the ($a$) panel of Fig.~\ref{fig:rotation}, although revolution is not necessary for the argument afterwards.
For more familiar example, we assume that boosted dark matter comes from the Sun~\cite{Berger:2014sqa,Kong:2014mia,Kachulis:2017nci}. The rotation axis is inclined by angle $\alpha$ with respect to the revolution axis, and more importantly, inclined by angle $\beta$ from the source direction. For position ($i$) whose $\delta$ is defined as 0, $\beta$ is assumed largest, i.e., winter solstice.
As the Earth moves around, $\beta$ decreases and becomes $90^{\circ}$ at position ($ii$) (i.e., $\delta=90^{\circ}$ and spring equinox), and finally is minimized at position ($iii$) (i.e., $\delta=180^{\circ}$ and summer solstice).
In this specific example, $\beta$ spans $66.6^{\circ}$ to $113.4^{\circ}$.
Here the detector location is described by polar angle $\gamma$ measured from the rotation axis, and the detector in Fig.~\ref{fig:rotation} is located in the northern hemisphere.

We now calculate analytically how many hours the detector at $\gamma$ benefits from the ``Earth Shielding'', assuming $\phi_{\rm cr}$ in Eq.~\eqref{eq:cr} is zero and the source of interest is point-like for simplicity.
For full generality, we consider an arbitrary $\delta$.
We let the rotation axis be along the $\hat{z}$ direction, and place $\hat{z}$, $\hat{y}$, and the signal direction vector in a common plane.
Given this geometrical configuration, it is easy to see that $\alpha'$, the joining angle between the revolution axis and the rotation axis, is given by
\bea
\sin\alpha' = \sin \alpha \cos \delta\,.
\eea
As shown in the ($b$) panel of Fig.~\ref{fig:rotation}, the detector of interest sweeps through along the red circle as the Earth rotates. An arbitrary coordinate on this circle $P$ is
\bea
P = \left(\sin\gamma \cos\varphi,\, \sin\gamma \sin \varphi,\, \cos\gamma \right)\,,
\eea
where $\varphi$ defines the azimuthal angle around the rotation axis. Here we employed a unit sphere as the radius is irrelevant.
On the other hand, the black circle defines the boundary dividing the region to which the ``Earth Shielding'' is applicable. An arbitrary coordinate along this circle, $Q$ is given by
\bea
Q = (\sin\psi,\, -\sin\alpha' \cos\psi,\, \cos\alpha' \cos\psi)\,.
\eea
Here we first parameterize the black circle on the $z$-$x$ plane in terms of a polar angle $\psi$ and rotate it about the $\hat{x}$ axis by $\alpha'$.
Obviously, $P=Q$ determines the two intersecting points $P_1$ an $P_2$ from which fractional day $D_f$ is calculated. We find that $D_f$ is
\bea
D_f = \left\{
\begin{array}{l l}
1 &\hspace{0.5cm}{\rm for }~\tan\gamma < \tan\alpha'\,, \\ [2mm]
0 &\hspace{0.5cm}{\rm for }~\tan\gamma < -\tan\alpha'\,, \\ [2mm]
\frac{2\sin^{-1}(-\tan\alpha'/\tan\gamma)-\pi}{2\pi} & \hspace{0.5cm}{\rm otherwise,}
\end{array}
\right. \label{eq:Hform}
\eea
where $\sin^{-1}(-\tan\alpha'/\tan\gamma) \in [\pi/2,\, 3\pi/2]$.
Eq.~\eqref{eq:Hform} implies that $D_f$ differs day-by-day because $\delta$ changes as the Earth revolves.
However, we observe the following identity,
\bea
D_f(\delta)+D_f(\delta+\pi)= \frac{2\sin^{-1}(-\tan\alpha'/\tan\gamma)-\pi}{2\pi}+\frac{2\sin^{-1}(\tan\alpha'/\tan\gamma)-\pi}{2\pi}= 1\,,
\eea
using the relation $\sin^{-1}(x)+\sin^{-1}(-x)=2\pi$ for any $x$.
This means that the detector is exposed to the signal {\it effectively} for half year every single revolution, which is consistent with one's intuition.

We next discuss how $D_f$ is affected by the Earth's rotation for the non-point-like source.
In particular, if a cut of sizable $\theta_C$ is imposed, it is important to see whether $D_f$ in Eq.~\eqref{eq:Hform} is still available or some additional correction should be made.
To understand the situation, it is convenient to view the travel of a non-point-like source on the celestial sphere (see the ($c$) panel of Fig.~\ref{fig:rotation}).
The signal source is moving along the red circle as the Earth rotates, and when it is placed below the Earth surface (green areas), the ``Earth Shielding'' is effective.
Three example instants are presented in the same panel. At instant A, the source core intersects the horizon, and thus a half cone (green area) benefits from the ``Earth Shielding'' while the other half cone (blue area) does not.
When it comes to instant B, the source core is located above the horizon. Nevertheless, some fraction of the $\theta_C$ cone benefits from the ``Earth Shielding''.
On the other hand, although the source core is below the horizon at instant C, some small fraction of the $\theta_C$ cone (blue area) has no benefit from the ``Earth Shielding''.
However, we see that the loss at instant C is compensated by the gain at instant B.
Therefore, we are allowed to collapse the source area of interest to a single point and use Eq.~\eqref{eq:Hform} as the corresponding $D_f$.

The argument that we have developed so far can be readily applied to the case where the signal is from the GC.
Unlike the case of the Sun, the Earth negligibly moves around the GC.
The situation is similar to one-day exposure in the case of the Sun.
The angle between celestial north and Galactic north is $62.87^{\circ}$, i.e., $\alpha = 62.87^{\circ}$, with respect to the GC and we estimate our solar system is located at $\delta=57.8^\circ$ from Eq.~(\ref{eq:Hform}).
Our benchmark detectors mentioned in Sec.~\ref{sec:surface} are or will be at Fermi National Accelerator Laboratory or CERN whose latitudes are $41.8^\circ$ and $46.3^\circ$, or equivalently, $\gamma_{\rm FNAL}=48.2$ and $\gamma_{\rm CERN}=43.7$, respectively.
Plugging angle values, we find that $D_f$'s for SBN detectors and ProtoDUNE detectors are 0.66 and 0.69, correspondingly.

\section{Phenomenology \label{sec:pheno}}

Armed with the experimental strategy, ``Earth Shielding'' delineated in the previous section, we devote this section to discussions on several applications.
We begin with enumerating applicable physics scenarios followed by reviewing briefly our benchmark DM model, two-component BDM scenario, to apply our proposal, and then exhibit phenomenologically intriguing example analyses relevant to the benchmark model.

\subsection{Applicable Scenarios \label{sec:scenarios}}

We recall that the threshold energies of (most) surface-based detectors enumerated in Table~\ref{tab:sumdetector} are $\mathcal O (10\,{\rm MeV})$.
Therefore, the most promising signals arise from the scattering of {\it boosted} SM-sector or dark-sector objects transferring energy above these thresholds to the target.
An example for the former category is the conventional DM annihilation into a neutrino-antineutrino pair ($\nu \bar \nu$).
Such neutrino events have been studied in various literature~\cite{Frankiewicz:2015zma, Aartsen:2015xej, Aartsen:2016exj, Aartsen:2016pfc} as an indirect signal of DM, coming from the GC or the Sun,
at the underground neutrino experiments as stated in the previous section.
We leave a dedicated study for this scenario to a future project~\cite{future} and focus on the other new possibility.
One of the representative examples belonging to the latter category, which we shall take as our benchmark DM scenario, is the relativistically produced lighter DM in models of two-component DM~\cite{Agashe:2014yua, Belanger:2011ww}.
We will show that various phenomenological studies with this benchmark scenario benefit from the ``Earth Shielding'' even though we are forced to sacrifice some portion of signal data.

\begin{figure}[t]
\centering
\includegraphics[width=8cm]{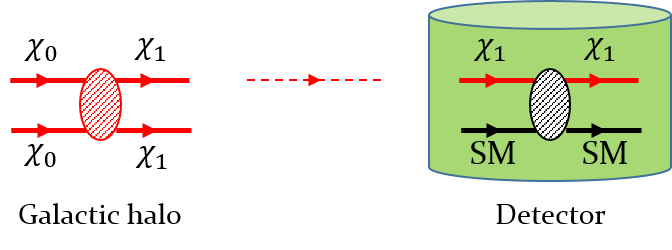}
\vspace{-0.2cm}
\caption{\label{fig:scenario} The minimal BDM scenario and elastic scattering of boosted (lighter) DM at a detector. }
\end{figure}

Let us give a concise review on essential features of the minimal BDM scenario.
For the two-component DM models,\footnote{The stability of the two DM species is often ensured by separate symmetries such as $Z_2 \otimes Z_2'$ and $U(1)' \otimes U(1)''$, e.g. the model in Ref.~\cite{Belanger:2011ww}.} the heavier one, say $\chi_0$, has no direct interaction with the SM sector but only through the lighter one, say $\chi_1$.
The assisted freeze-out mechanism~\cite{Belanger:2011ww} dictates the observed relic abundance, which is dominated by the heavier DM component in typical cases, and the cosmological and astrophysical features are mostly the same as those in conventional scenario of weakly interacting massive particle (WIMP).
As shown in Fig.~\ref{fig:scenario}, in the present universe, $\chi_0$ can pair-annihilate into a pair of $\chi_1$'s.
We are interested in the scenario where the $\chi_1$ comes with a large boost factor due to the large difference between the $\chi_0$ mass, $m_0$, and the $\chi_1$ mass, $m_1$~\cite{Agashe:2014yua}.
Note that this is not the only mechanism of having boosted or, at least, fast-moving DM in the universe today; for example, one can consider models with a $Z_3$ symmetry which give rise to the semi-annihilation process~\cite{DEramo:2010keq} or models involving anti-baryon numbered DM-induced nucleon decays inside the Sun although the produced DM is not so energetic~\cite{Huang:2013xfa}.

A viable way to probe two-component (in general, multi-component) BDM scenarios is to observe experimental signatures induced by the relativistic $\chi_1$ coming from an area in which $\chi_0$ density is high enough.
Examples include the GC~\cite{Agashe:2014yua, Necib:2016aez, Alhazmi:2016qcs, Kim:2016zjx, Giudice:2017zke}, the Sun~\cite{Berger:2014sqa, Kong:2014mia, Alhazmi:2016qcs}, and dSphs~\cite{Necib:2016aez}.
Such boosted $\chi_1$ can scatter off detector material in large-volume neutrino experiments~\cite{Agashe:2014yua, Necib:2016aez, Alhazmi:2016qcs, Kim:2016zjx, Berger:2014sqa, Kong:2014mia} or conventional WIMP direct detection experiments~\cite{Giudice:2017zke}, and signal detection rates depend on the flux of $\chi_1$, the incoming energy of $\chi_1$, and/or relevant detection threshold energy.  All of them are operated deep underground in order to control enormous cosmic-ray backgrounds.
We point out that for scenarios accompanying additional (secondary) processes on top of the primary target recoil, e.g., the so-called inelastic BDM, hence allowing for the separation of a variety of background events~\cite{Kim:2016zjx, Giudice:2017zke}, it is possible to probe the signals even in surface-based experiments such as ProtoDUNE~\cite{Chatterjee:2018mej}.
However, the elastic scattering of $\chi_1$ usually suffers from uncontrollable cosmic-ray backgrounds at surface detectors.
Therefore, the search for the elastic BDM signature necessitates utilizing the ``Earth Shielding''.

\subsection{Analysis Details  \label{sec:analysis}}

We are now in the position to discuss phenomenology of the aforementioned two-component BDM scenario at surface-based experiments.
Experimental sensitivities are generically obtained by the number of signal events excluded at, for example, 90\% C.L.
The expected number of signal $N_{\rm Sig}$ is given by
\begin{align}
N_{\rm Sig} = \sigma_\epsilon \; D_f \;\mathcal F \;  t_{\rm exp} \; N_T\, , \label{eq:Nsig}
\end{align}
where $T$ stands for the target that $\chi_1$ scatters off, $\sigma_\epsilon$ is the cross section for the process $\chi_1 T \to \chi_1 T$, ${\mathcal F}$ symbolizes the flux of $\chi_1$, $t_{\rm exp}$ denotes the exposure time, and $N_T$ implies the number of target particles in the detector fiducial volume $V_{\rm fid}$ of interest.
We factored out $D_f$, so $\mathcal{F}$ is the genuine flux of $\chi_1$ for a given $t_{\rm exp}$.
One should note that in our notation $\sigma_\epsilon$ includes realistic effects such as the acceptance from cuts, threshold energy, and detector response, hence it can be understood as the fiducial cross section.

The 90\% C.L. exclusion limit $N^{90}$ is calculated with a modified frequentist construction~\cite{cls1,cls2}. Here we follow the method in Refs.~\cite{Dermisek:2013cxa,Dermisek:2014qca} where the Poisson likelihood is assumed.
An experiment is said to be sensitive to a given signal in a {\it model-independent} fashion if $N_{\rm Sig} \ge N^{90}$.
Substituting Eq.~\eqref{eq:Nsig} into this inequality, we find
\begin{align}
\sigma_\epsilon \mathcal F \ge \frac{N^{90}}{D_f \, t_{\rm exp}\, N_T}\,.
\label{eq:sensitivity}
\end{align}
All the quantities in the right-hand-side of~\eqref{eq:sensitivity} are constant, and fully determined once a target experiment, hence the associated detectors are chosen.
On the other hand, the flux of $\chi_1$ under a BDM scenario in the left-hand side of~\eqref{eq:sensitivity} is a function of the $\chi_0$ mass $m_0$:
\bea
\mathcal F &=& \frac{1}{2}\cdot \frac{1}{4\pi}\int d\Omega \int_{\rm los}ds \langle \sigma v \rangle_{\chi_0 \overline{\chi}_0 \to \chi_1 \overline{\chi}_1}\left(\frac{\rho(s,\theta)}{m_0} \right)^2 \nonumber \\
&=& 1.6 \times 10^{-4}\,{\rm cm}^{-2} {\rm s}^{-1}
\, \times \, \left( \frac{\langle \sigma v \rangle_{\chi_0 \overline{\chi}_0 \to \chi_1 \overline{\chi}_1}}{5 \times 10^{-26}\,{\rm cm}^3 {\rm s}^{-1}} \right) \times \left( \frac{{\rm GeV}}{m_0}\, \right)^2  \label{eq:fluxformula}\\
&\equiv& \mathcal{F}_{\rm ref}^{\rm 180^\circ} \times \left( \frac{\langle \sigma v \rangle_{\chi_0 \overline{\chi}_0 \to \chi_1 \overline{\chi}_1}}{5 \times 10^{-26}\,{\rm cm}^3 {\rm s}^{-1}} \right) \times \left( \frac{{\rm GeV}}{m_0} \right)^2 \,, \nonumber
\eea
where $\rho$ describes the $\chi_0$ density distribution in terms of the line-of-sight $s$ and solid angle $\Omega$, $\mathcal{F}_{\rm ref}^{\rm 180^\circ}$ denotes the reference flux value in the second line, and $\langle \sigma v \rangle_{\chi_0 \overline{\chi}_0 \to \chi_1 \overline{\chi}_1}$ is the velocity-averaged annihilation cross section of $\chi_0 \overline{\chi}_0 \to \chi_1 \overline{\chi}_1$ at the present universe.
Note that we implicitly assume $\chi_0$ and its anti-particle $\overline{\chi}_0$ are distinguishable.
For the indistinguishable case, one should simply remove the overall prefactor 1/2 and repeat the calculation, accordingly.

\begin{figure}[t]
\centering
\includegraphics[width=8.4cm]{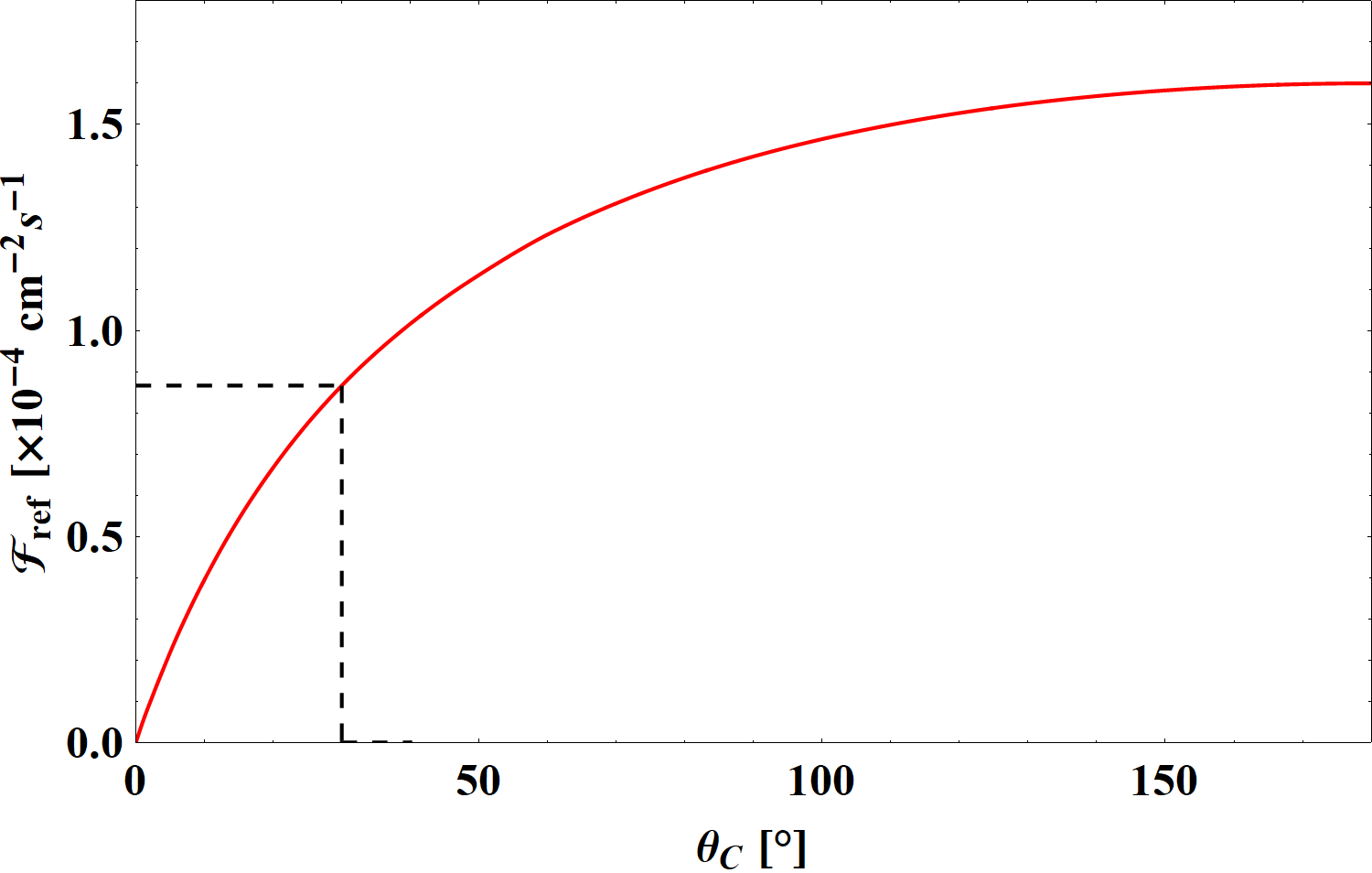}
\caption{\label{fig:fluxvsangle} $\mathcal{F}_{\rm ref}$ as a function of $\theta_C$. $\mathcal{F}_{\rm ref}^{\rm 180^\circ}$ is normalized to $1.6 \times 10^{-4}\,{\rm cm}^{-2} {\rm s}^{-1}$ together with the NFW DM halo profile~\cite{Navarro:1995iw,Navarro:1996gj}.
}
\end{figure}

Here to find $\mathcal{F}_{\rm ref}^{\rm 180^\circ}$, we average out the total flux per unit time over the whole sky assuming NFW DM density profile~\cite{Navarro:1995iw,Navarro:1996gj} for $\rho(s,\theta)$ with local DM density $\rho_\odot = 0.3\,{\rm GeV}/{\rm cm}^3$ near the Sun which is 8.33 kpc away from the GC, scale radius $r_s = 24.42$ kpc, scale density $\rho_s = 0.184\,{\rm GeV/cm}^3$, slope parameter $\gamma = 1$, $m_0=1$ GeV, and $\langle \sigma v \rangle_{\chi_0 \overline{\chi}_0 \to \chi_1 \overline{\chi}_1}=5 \times 10^{-26}\,{\rm cm}^3 {\rm s}^{-1}$.
The chosen value for the present-day velocity-averaged cross section is consistent with the observed DM abundance.
This is true for BDM scenarios where the dominant relic density is determined by the $s$-wave annihilation process $\chi_0 \overline{\chi}_0 \to \chi_1 \overline{\chi}_1$.
A restriction on the signal region may affect the reference flux value; for example, if we integrate the integrand in the first line of Eq.~\eqref{eq:fluxformula} (i.e., the differential flux) only within a $30^\circ$ cone around the GC for the BDM scenario, it becomes reduced by a factor of 2, i.e., $2 \mathcal{F}_{\rm ref}^{30^\circ} \approx \mathcal{F}_{\rm ref}^{\rm 180^\circ}$ (see the black dashed lines in Fig.~\ref{fig:fluxvsangle}).
Fig.~\ref{fig:fluxvsangle} shows $\mathcal{F}_{\rm ref}$ as a function of $\theta_C$ for the ordinary BDM scenario, which can be obtained by integrating $\rho$ in the first line of Eq.~\eqref{eq:fluxformula} over $s$ and $\theta\in[0, \theta_C]$.

The experimental sensitivity in~\eqref{eq:sensitivity} now becomes a more familiar form,
\bea
\sigma_\epsilon  \ge \frac{N^{90}}{D_f \, t_{\rm exp} \, N_T \, \mathcal{F}_{\rm ref}^{\theta_C}}\left( \frac{5 \times 10^{-26}\,{\rm cm}^3 {\rm s}^{-1}}{\langle \sigma v \rangle_{\chi_0 \overline{\chi}_0 \to \chi_1 \overline{\chi}_1}} \right)\left( \frac{m_0}{{\rm GeV}} \right)^2 \,, \label{eq:forBDM}
\eea
which is reminiscent of a well-known parameterization of spin-independent or spin-dependent WIMP-nucleon scattering cross section versus WIMP mass in standard WIMP direct searches.
For the BDM scenarios, all model-dependent information associated with the coupling of $\chi_1$ to SM particles is encoded in $\sigma_\epsilon$, so that the experimental sensitivity determined by the right-hand side of~\eqref{eq:forBDM} does {\it not} depend on modeling of $\chi_1-{\rm SM}$ interactions.
By contrast, $\langle \sigma v \rangle_{\chi_0 \overline{\chi}_0 \to \chi_1 \overline{\chi}_1}$ is deeply related to the cosmological history of the universe, and $\mathcal{F}_{\rm ref}^{\theta_C}$ encapsulates information on the DM halo distribution.
Once a specific cosmology and a DM halo model are chosen, then the right-hand side of~\eqref{eq:forBDM} essentially becomes proportional to $m_0^2$ and constant factors determined by detector characteristics.
Obviously, weaker-coupled regions (up to acceptance and efficiency) can be probed towards larger $t_{\rm exp}$, larger $N_T$, and smaller $m_0$.

\begin{table}[t]
\centering
\begin{tabular}{c | c c | c c}
\hline
 \multirow{2}{*}{Detector} & \multicolumn{2}{c|}{$N^{90}$} & \multicolumn{2}{c}{$N_{\rm BG}$} \\
 \cline{2-5}
 & \hspace{0.2cm} All sky \hspace{0.2cm} & \hspace{0.2cm} $30^\circ$ \hspace{0.2cm} & \hspace{0.2cm} All sky \hspace{0.2cm} & \hspace{0.2cm} $30^\circ$ \\
 \hline \hline
ProtoDUNE-DP & 5.43 & 2.77 & 5.82 & 0.39 \\
ProtoDUNE-SP & 6.18 & 2.93 & 8.31 &  0.56 \\
\hspace{0.2cm} ProtoDUNE-total\hspace{0.2cm} & 7.59 & 3.23 & 14.1 & 0.95 \\
\hline
MicroBooNE & 3.57 & 2.45 & 1.46 & 0.098 \\
SBND & 3.80 & 2.48 & 1.85 & 0.12 \\
ICARUS & 6.08 & 2.91 & 7.95 & 0.53 \\
SBN Program-total & 6.94 & 3.09 & 11.3 & 0.75 \\
\hline
\end{tabular}
\caption{\label{table:n90}
$N^{90}$ values for all sky (second column) and $30^\circ$ (third column) at various surface detectors.
The numbers are computed with respect to the expected number of (atmospheric neutrino-induced) background events ($N_{\rm BG}$) during one-year operation reflecting the corresponding $D_f$.
}
\end{table}

\begin{figure}[t]
\centering
\includegraphics[width=7cm]{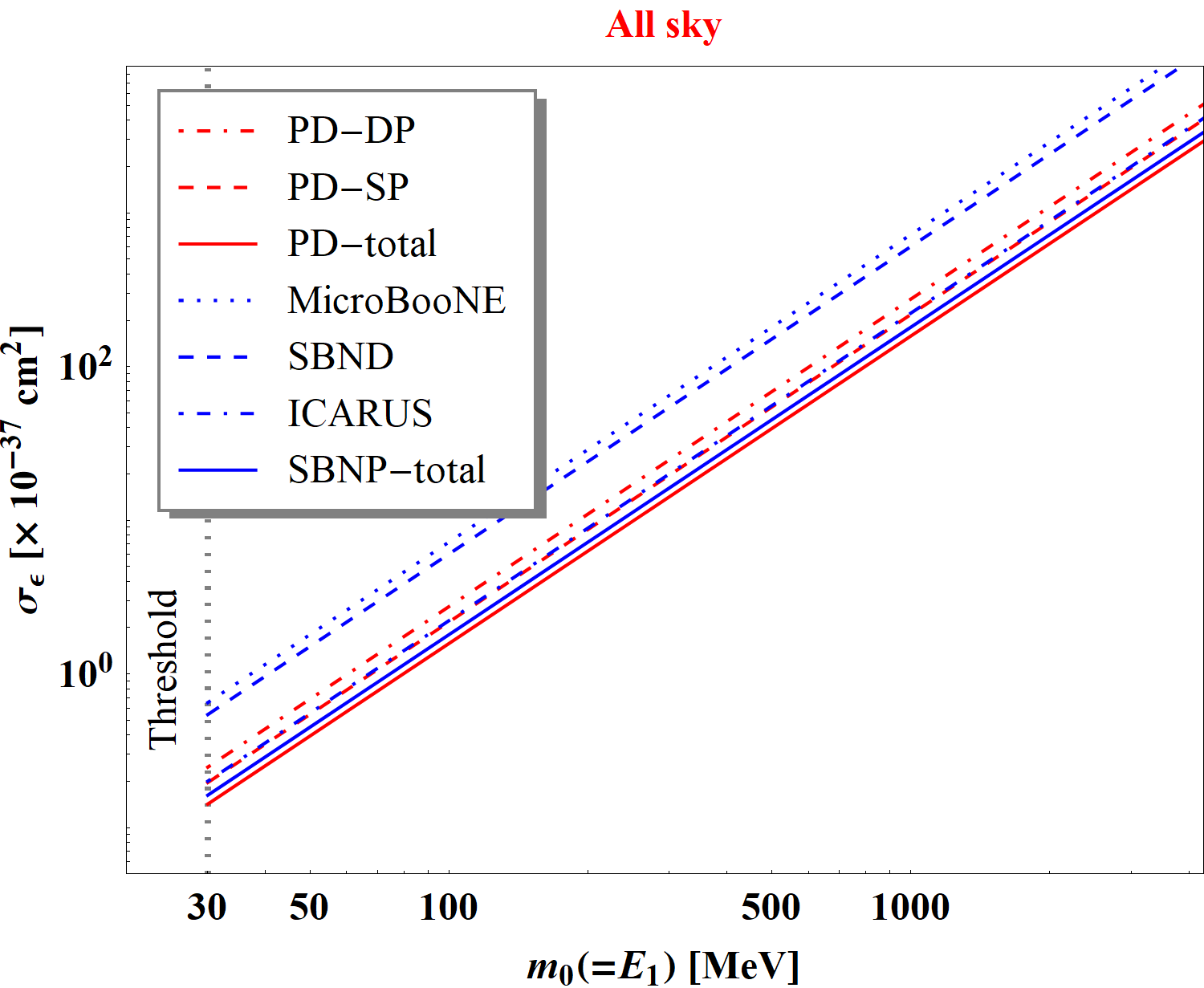} \hspace{0.2cm}
\includegraphics[width=7cm]{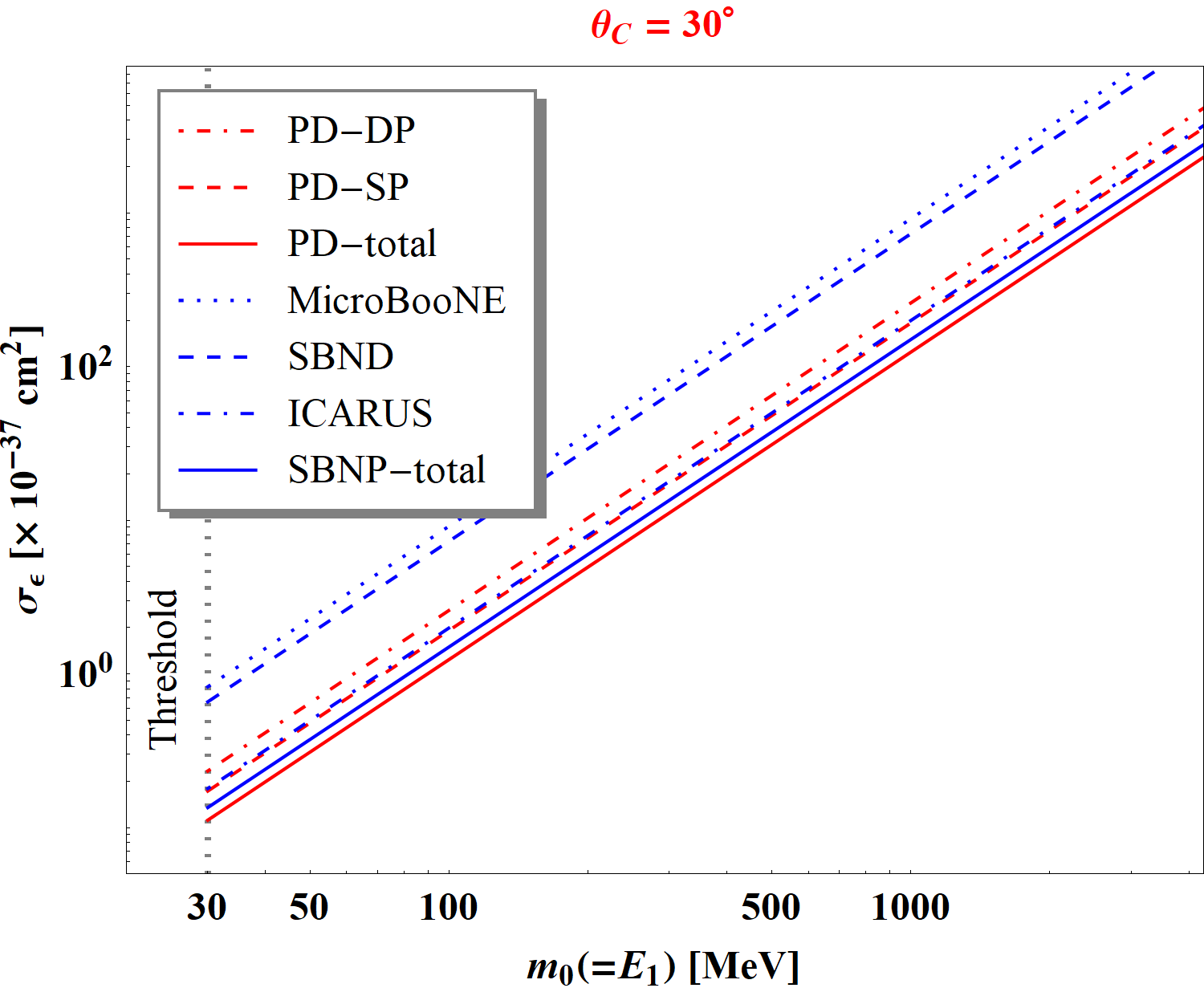}
\caption{\label{fig:sensitivity}
The 90\% C.L. experimental sensitivities to the elastic BDM signal via electron scattering at the detectors listed in Table \ref{table:n90}, together with $\langle \sigma v \rangle_{\chi_0 \overline{\chi}_0 \to \chi_1 \overline{\chi}_1} = 5\times 10^{-26}\,{\rm cm}^3 {\rm s}^{-1}$ (i.e., annihilation scenario of BDM).
Assumed is one-year exposure from all sky (left panel) and a $30^\circ$ cone (right panel). The vertical black dotted lines denote the threshold energy for electrons. We take a conservative value of 30 MeV for SBN Program detectors.
}
\end{figure}

In Fig.~\ref{fig:sensitivity}, we exhibit the experimental sensitivities with respect to the BDM searches at ProtoDUNE SP (red dotdashed), ProtoDUNE DP (red dashed), ProtoDUNE-total (red solid), MicroBooNE (blue dotted), SBND (blue dashed), ICARUS (blue dotdashed), and SBN Program-total (blue solid).\footnote{We henceforth call ProtoDUNE and SBN Program as PD and SBNP, respectively, when mentioning them in the figure legends to save the space for graphical objects.}
For illustration, we consider elastic scattering of boosted $\chi_1$ off electrons in the detector material, but similar analyses for proton target can be readily carried out.
Two signal regions are analyzed -- 1) all sky shown in the left panel and 2) $30^\circ$ cone shown in the right panel -- together with $\langle \sigma v \rangle_{\chi_0 \overline{\chi}_0 \to \chi_1 \overline{\chi}_1} = 5\times 10^{-26}\,{\rm cm}^3 {\rm s}^{-1}$ (i.e., annihilation scenario of BDM).
The above-chosen $30^\circ$ of angle cut allows (almost) the best 90\% C.L. limits for ProtoDUNE-total and SBN Program-total, and we observe an improvement by 10\% orders of magnitude, in the experimental sensitivities for ProtoDUNE-total and SBN Program-total, comparing all sky and $30^\circ$-cone results in Fig.~\ref{fig:sensitivity}.
We further assume one-year data collection, for which the expected number of neutrino-induced backgrounds for each detector straightforwardly follows from the detector fiducial volume tabulated in Table~\ref{tab:sumdetector} and Eq.~\eqref{eq:Nallsky}.
Again be aware that only the events coming from the detector bottom surface are analyzed, so we effectively consider $D_f$-year exposure which are already taken into account in Eq.~\eqref{eq:Nallsky}, i.e., 3600 s\,hr$^{-1}\times$ 24 hr\,day$^{-1}\times$ 365 day $\times \, D_f \approx 2.18 \times 10^7$ s and $2.08 \times 10^7$ s for ProtoDUNE and SBN Program, respectively.
We compute the corresponding $N^{90}$ and $N_{\rm BG}$ values accordingly and collect them in Table~\ref{table:n90} just for convenience.

Several comments follow in order.
First of all, we observe that the choice of $\theta_C=30^\circ$ is not optimal for MicroBooNE and SBND because their respective $t_{\rm exp} N_T$ is too small so that an almost background-free environment is kept even with no $\theta_C$ cut (i.e., $N_{\rm bkg}^{180^\circ} \lesssim 1.1~{\rm yr}^{-1}$).
In general, for any data accumulation with very small $t_{\rm exp} N_T$, increasing $\theta_C$ ensures better experimental sensitivities.
Secondly, we note that the overall flux $\mathcal{F}$ contains information on the production mechanism of boosted $\chi_1$, i.e., annihilation versus decay.
Although our choice in this paper is an annihilation model $\chi_0 \overline{\chi}_0 \to \chi_1 \overline{\chi}_1$, the result is directly applicable to semi-annihilation models~\cite{DEramo:2010keq}, e.g., $\chi_0 \chi_0 \to \chi_0 \phi$ with $\phi$ being either a SM or dark-sector particle as two $\chi_0$'s are involved in the initial state in both classes of models.
On the contrary, decay models, e.g., $\chi_0 \to \chi_1 \overline{\chi}_1$ (or with additional radiation~\cite{Kopp:2015bfa,Bhattacharya:2014yha, Bhattacharya:2016tma}) involves a single $\chi_0$ in the initial state so that each sensitivity plot would have a slope half that of the corresponding plot for the annihilation model with a constant shift in the $\sigma_\epsilon$ direction and $\langle \sigma v \rangle$ replaced by the decay width of $\chi_0$, with the optimal angle cut being all sky as shown in the right panel of Fig.~\ref{fig:Sig-Compare}.
Thirdly, we remark that $\mathcal{F}_{\rm ref}^{\theta_C}$ carries the $\chi_0$ halo model dependence, as mentioned earlier. Imposing a nontrivial $\theta_C$ cut, we start to see some $\chi_0$ halo model-specific dependence relative to the corresponding $\mathcal{F}_{\rm ref}^{180^\circ}$.
In this sense, the experimental sensitivity resulting from all-sky data provides a halo-independent limit up to the total flux.
A larger (smaller) total flux leads a constant shift of the lines downward (upward).
Fourthly, any electron recoil should exceed the electron threshold energy $E_{\rm th}$, so the kinematically accessible minimum $m_0$ is the same as $E_{\rm th}$.
For SBN Program detectors, we conservatively take the same magnitude of $E_{\rm th}$ as that for ProtoDUNE detectors, denoting them commonly by black dotted vertical lines.
Finally, direction information is inferred from the recoil track, but they may differ slightly~\cite{Necib:2016aez}.
Here we take the simplified approach that the directionality of an event can be extracted reasonably well due to good angular resolution ($\theta_{\rm res}  \sim 1^\circ$ as in Table~\ref{tab:sumdetector}), good track reconstruction of LArTPC-based detectors, and large enough $\chi_1$ energy $E_1 >$ 30 MeV compared to the target mass $m_e=0.511$ MeV.

We can translate these model-independent experimental sensitivities into the search limits in terms of model parameters if the interaction between the $\chi_1$ and SM particles is specified.
In this paper, we pick a dark photon scenario for illustration in which the relevant Lagrangian terms are summarized to
\begin{align}
\mathcal L \supset -\frac{\epsilon}{2}F_{\mu\nu}X^{\mu\nu} + g_D \bar{\chi}_1\gamma^{\mu}\chi_1 X_\mu \,. \label{eq:lagrangian}
\end{align}
The first term describes the kinetic mixing between U(1)$_{\rm EM}$ and U(1)$_{\rm X}$ parameterized by the small number $\epsilon$.
$F_{\mu \nu}$ and $X_{\mu \nu}$ are the field strength tensors for the ordinary photon and the dark photon, respectively.
The second term with the associated interaction strength parameterized by $g_D$ determines the coupling of the dark sector to the SM sector, mediated by the dark photon $X_\mu$.

\begin{figure}[t]
\begin{center}
\includegraphics[width=7cm]{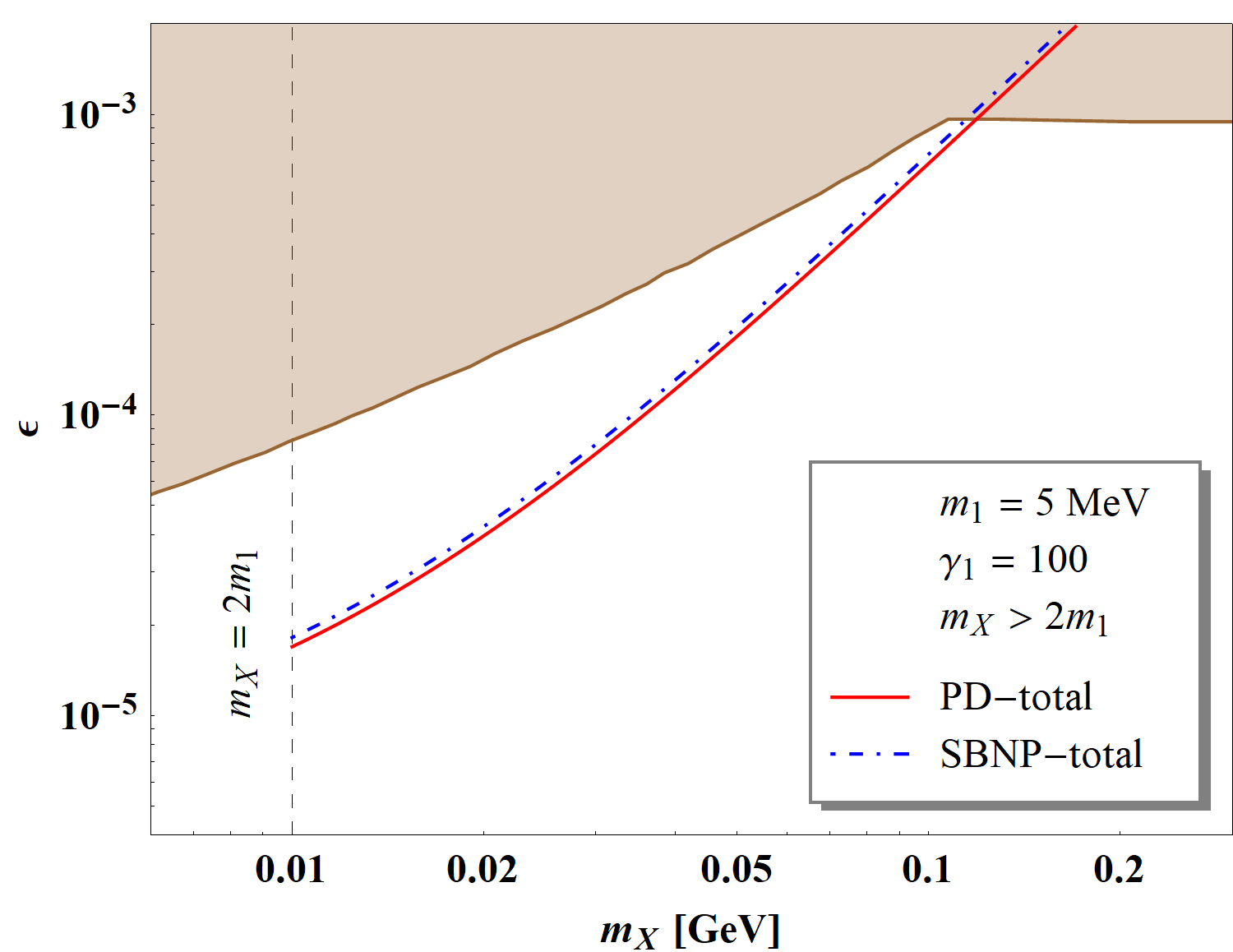} \hspace{0.2cm}
\includegraphics[width=7cm]{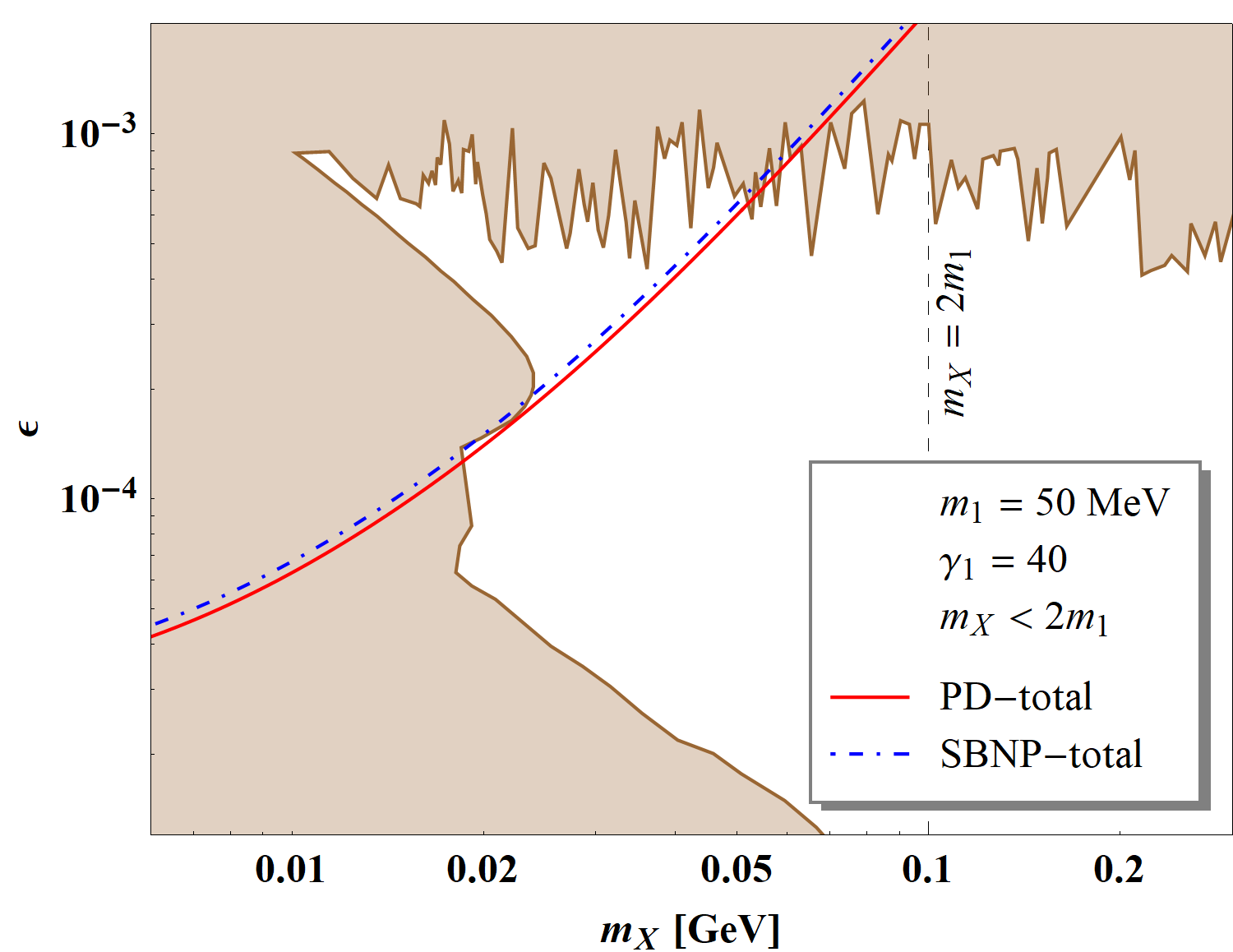}
\caption{\label{fig:dp}
Experimental sensitivities in the dark photon model parameters $m_X - \epsilon$.
Our limits are given by BDM searches in the elastic electron scattering channel arising in the benchmark model described in~\eqref{eq:lagrangian}.
The left  (right) panel exhibits the result for the case where the dark photon $X$ decays invisibly (visibly), with one-year data collection from all sky.
The brown-shaded regions show the current excluded parameter space by 90\% C.L., according to the reports in Refs.~\cite{Banerjee:2017hhz} (left panel) and~\cite{Banerjee:2018vgk} (right panel). For both cases, the dark-sector gauge coupling $g$ is set to be unity for simplicity.
}
\end{center}
\end{figure}

In the left (right) panel of Fig.~\ref{fig:dp}, we show the experimental sensitivity in terms of the dark photon mass and the kinetic mixing parameter, i.e., $m_X - \epsilon$, for the case where the dark photon predominantly decay invisibly (visibly).
Assuming one-year (effectively, $D_f$-year) data collection from the whole sky, we consider the total combinations of detectors at ProtoDUNE (red solid curves) and SBN Program (blue dot-dashed curves).
To find the boundary values along the curves, we fix $m_1$, $m_0$ (or, equivalently, $\gamma_1 m_1$ for the pair-annihilation of non-relativistic $\chi_0$ to a $\chi_1$ pair), and the dark-sector gauge coupling (taking $g_D=1$), followed by numerically computing the $\epsilon$ value for a given $m_X$ to yield the associated $N^{90}$ signal events with the $E_{\rm th}$ for electron taken into account.
Note that the brown-colored areas denote the current ruled-out parameter regions by 90\% C.L. whose boundary values are obtained in Refs.~\cite{Banerjee:2017hhz} (left panel) and~\cite{Banerjee:2018vgk} (right panel).
We see that searches in the elastic electron scattering channel enable to probe parameter regions unexplored by past experiments with about an order of magnitude better sensitivity.
Here ProtoDUNE-total and SBN Program-total show similar parameter reaches due to their comparable fiducial volumes.

\begin{figure}[t!]
\centering
\includegraphics[width=6.7cm]{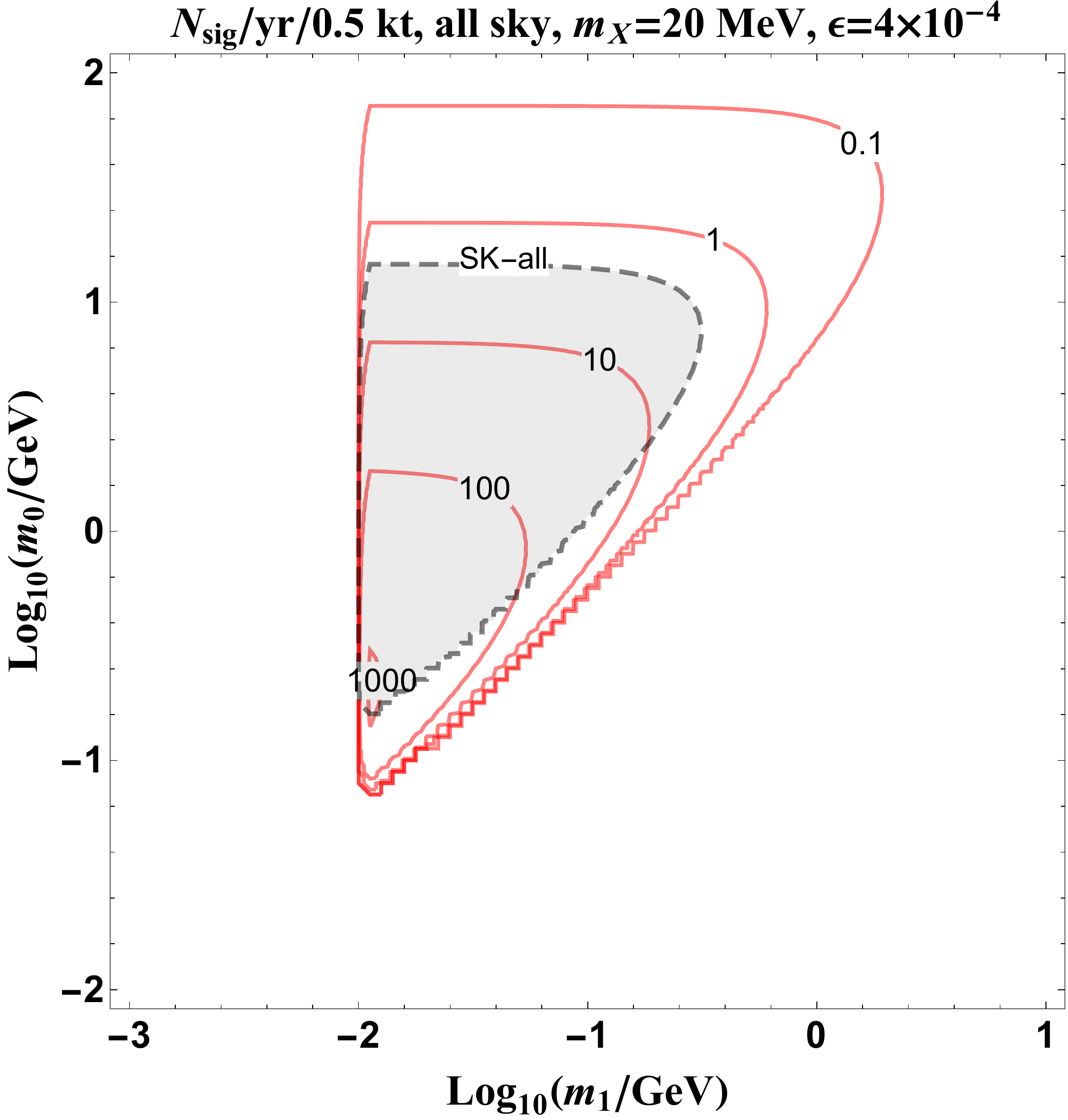}\hspace{0.5cm}
\includegraphics[width=6.7cm]{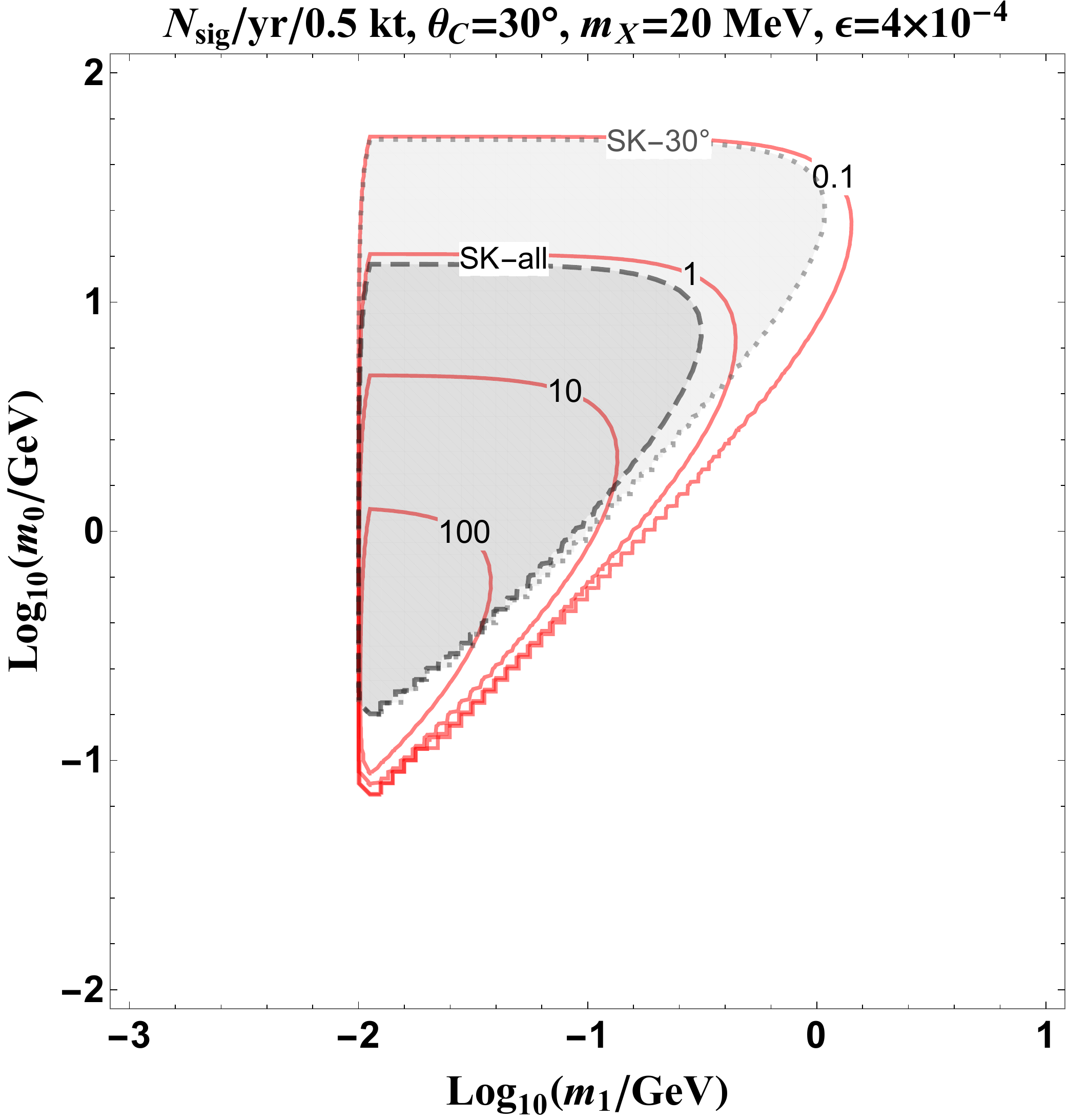}
\caption{\label{fig:Nsignal} The number of signal events per year for all sky (left panel) and 30$^\circ$ cone (right panel) with $V_{\rm fid} = 0.5$ kt, $m_X=20$ MeV, and $g_D \cdot\epsilon = 4\times 10^{-4}$.
The darker gray-shaded area represents the 90\% exclusion bounds with all sky data from SK (13.6 yr)~\cite{Dziomba:2012paz, Richard:2015aua}, assuming 10\% systematic uncertainty in the estimation of background number of events.
The light-shaded area in the right panel represents current BDM bounds using 161.9 kt$\cdot$yr of data reported by SK~\cite{Kachulis:2017nci}.
}
\end{figure}

\begin{figure}[t]
\centering
\includegraphics[width=6.7cm]{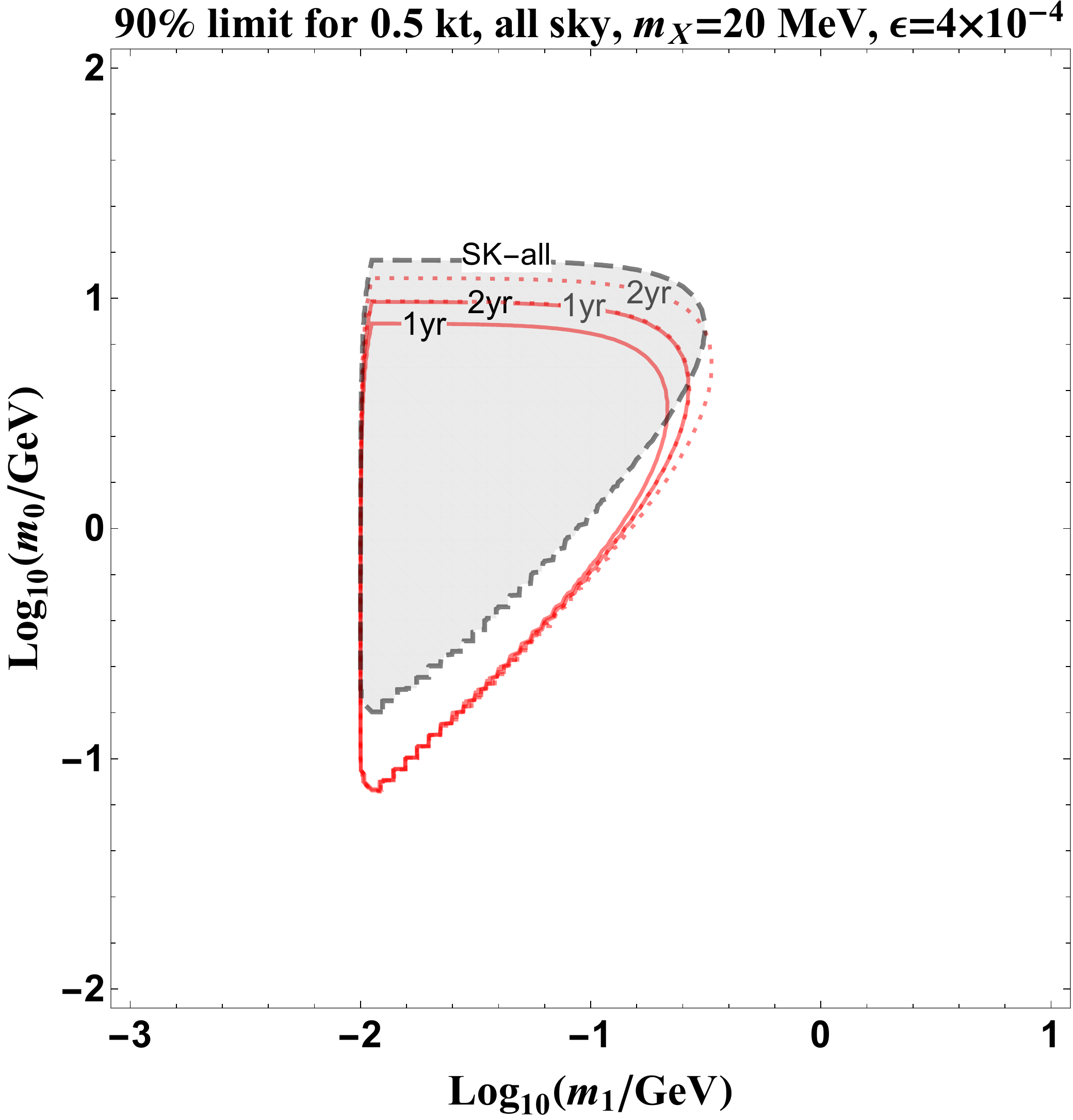}\hspace{0.5cm}
\includegraphics[width=6.7cm]{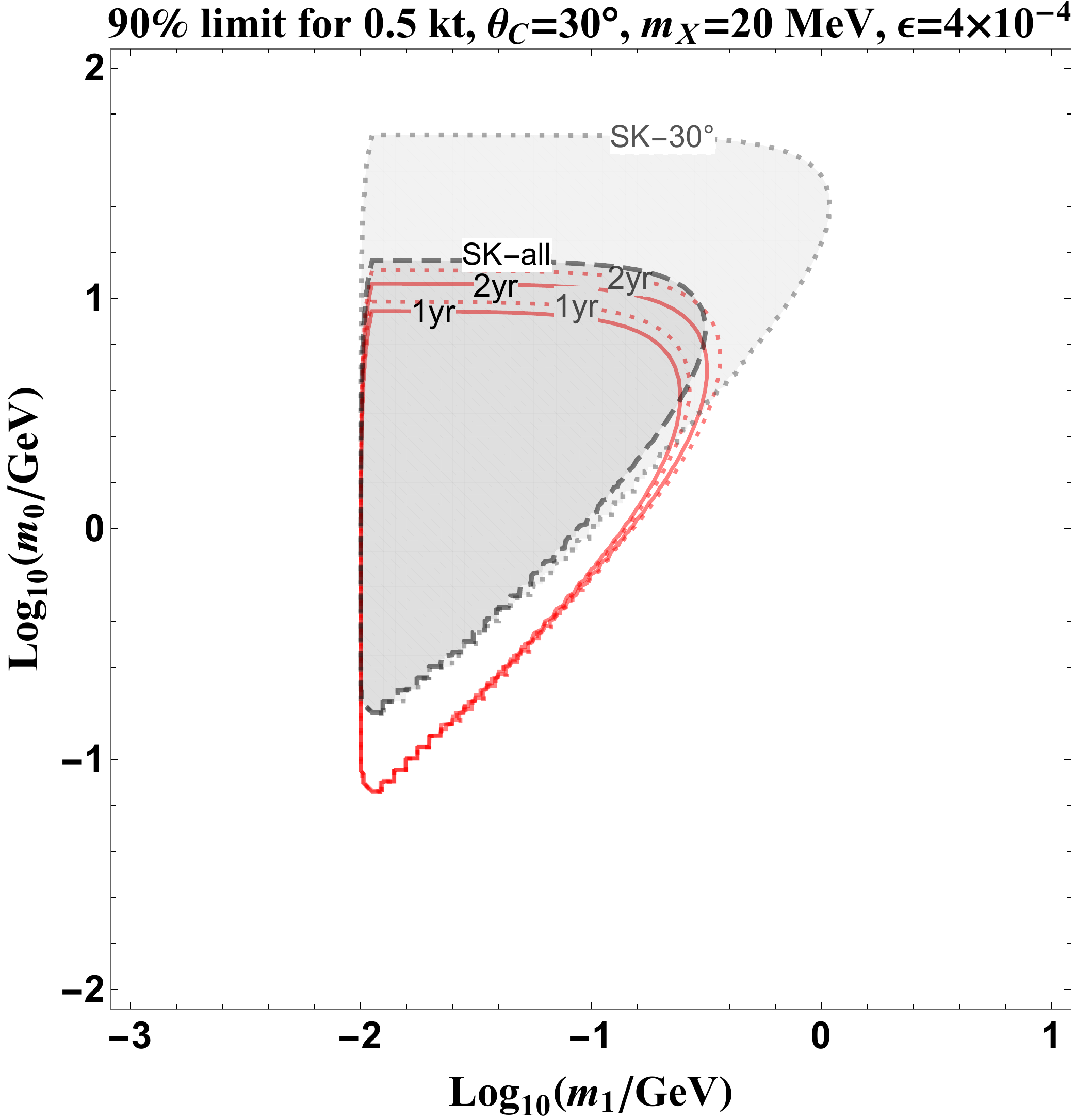}
\caption{\label{fig:exclusion}
The expected 90\% exclusion bounds from 1-year and 2-year running of 0.5 kt-$V_{\rm fid}$ detectors for all sky (left panel) and 30$^\circ$ cone (right panel).
The red-dotted curves are the same but using the improved estimation of neutrino backgrounds $N_{\rm bkg}^{180^\circ}=12.8$ yr$^{-1}$kt$^{-1}$ from Ref.~\cite{Necib:2016aez}.
The gray-shaded areas are bounds from SK measurements as in Fig.~\ref{fig:Nsignal}.
}
\end{figure}

While the previous analysis scheme constrains the sector relating $\chi_1$ with SM particles, one can interpret the same set of data, focusing on the sector connecting $\chi_0$ and $\chi_1$.
As the first example, in Fig.~\ref{fig:Nsignal}, we demonstrate the number of signal events per year (in red contours) for all-sky data (left panel) and 30$^\circ$-cone data (right panel) in the standard parameterization of $m_0$ versus $m_1$, fixing $V_{\rm fid} = 0.5$ kt, $m_X=20$ MeV, and $g_D \cdot \epsilon = 4\times 10^{-4}$.\footnote{
$D_f=0.69$ corresponding to CERN, the location of the ProtoDUNE detectors, is assumed.
}
Our choice of $m_X$ and $\epsilon$ is safe from current experimental bounds for $m_X < 2 m_1$~\cite{Banerjee:2017hhz}.
The dark gray-shaded area represents the 90\% C.L. exclusion bound inferred from the atmospheric neutrino measurement in Ref.~\cite{Dziomba:2012paz, Richard:2015aua} for which data over the whole sky was collected for 13.6 years by the SK Collaboration.
We provide the same limit from the SK all sky data for the 30$^\circ$-cone case since angular information of the data is not available.
We use the fully contained single-ring $e$-like events including both sub-GeV (0-decay electron events only) and multi-GeV as a conservative estimation of a total of 10.7 years~\cite{Dziomba:2012paz} and normalize the rate to 13.6 years~\cite{Richard:2015aua}.
Here we include 10\% systematic uncertainty in the estimation of background number of events.
On the other hand, the light-shaded area in the right panel represents current bounds from the BDM search in the elastic electron scattering channel using 161.9 kt$\cdot$yr of data observed by the SK Collaboration~\cite{Kachulis:2017nci}.
This recent analysis classifies the observed events in three different energy bins for several choices of angular cones ($\leq 40^\circ$), which greatly improves the exclusion limit, as clearly shown in the right panel.
The data corresponding to all sky is not available, so we only show the bound for 13.6 years in the left panel.
Obviously, LArTPC-based detectors allow to explore the parameter space towards the lower-right uncovered by SK.
This is essentially the region where the relevant electron recoil energy is lower than $\sim100$ MeV.

Similarly, the expected 90\% C.L. exclusion of 0.5 kt-$V_{\rm fid}$ detectors are shown as red contours in Fig.~\ref{fig:exclusion}, covering the diagonal boundary.\footnote{Remember that the fiducial volume for ProtoDUNE-total (SBN Program-total) is slightly above (below) 0.5 kt.}
Again, note that one-year (two-year) data effectively corresponds to $D_f$-year ($2D_f$-year) exposure with $D_f=0.69$.
A recent study \cite{Necib:2016aez} shows that an optimized analysis using GENIE neutrino Monte Carlo software can reduce the number of atmospheric neutrino background down to $N_{\rm bkg}^{180^\circ}=12.8$ yr$^{-1}$kt$^{-1}$.
This is about a factor of 3 reduction from the previous estimate as in~(\ref{eq:nuflux}).
With this improved estimation of background, we revisit the expected 90\% exclusion bounds from 1-year and 2-year running of 0.5 kt-$V_{\rm fid}$, and the corresponding results are shown in Fig.~\ref{fig:exclusion} as red-dotted curves for all sky (left panel) and 30$^\circ$ cone (right panel), respectively.
The improvement of the experimental sensitivities with this new background estimation is reduced for 30$^\circ$ cone compared to all sky since the signal-to-background ratio increases due to the reduced number of background events resulting in an optimal search angle $\theta_C > 30^\circ$ as discussed in Sec.~\ref{Earth Shielding}.


\section{Conclusions and Outlook \label{sec:conclusions}}

In general, it is challenging to fulfill physics analyses for rare signals coming from the sky, which are observed at surface-based detectors, because the number of cosmic-ray events, part of which potentially mimic the signal of interest, is huge.
In particular, if the expected experimental signature is featureless, separating the rare signal events out of cosmic-origin background ones is almost impossible.
In light of this situation, we claimed that a reasonable extent of signal isolation is nevertheless achievable at the price of a certain fraction of signal events.
The idea behind it is to restrict to the events coming out of the Earth surface so that potential cosmic-induced background events are significantly suppressed while they penetrate the Earth, being essentially left with neutrino-induced events.
The directionality information of an event is of crucial part, and therefore, the proposed method is readily applicable for the cases in which the signal of interest accompanies target recoil with a sizable track and the associated detector is good enough at the track measurement.

For the sake of validating our main idea, we considered the elastic scatterings of the lighter DM particles created by an annihilation/decay of the dominant DM relic, at several benchmark surface (or nearly surface) detectors including SBN Program and ProtoDUNE preceded by careful estimates for possible backgrounds.
We found that a sufficient level of signal sensitivities can be achieved, demonstrating the experimental sensitivities in several ways.
First of all, we exhibited (model-independent) experimental reaches at the benchmark detectors in the plane of (fiducial) signal cross section versus the mass of dominant DM, $m_0$ (see Fig.~\ref{fig:sensitivity}).
Since our BDM scenario under consideration involves a dark photon, which mediates the interaction between the lighter DM and the SM fermions, we plotted the possible coverage of parameter space in the $m_X-\epsilon$ plane as shown in Fig.~\ref{fig:dp}, from which we observed that some extent of uncovered parameter space can be probed at the benchmark detectors.
Finally, in Figs.~\ref{fig:Nsignal} and~\ref{fig:exclusion}, we translated the experimental reaches in terms of a standard model parameterization of BDM scenarios, i.e., $m_0$ versus $m_1$, and found that our benchmark detectors can access a large area of model space unexplored by the SK experiment~\cite{Kachulis:2017nci}.

In summary, our proposal with the ``Earth Shielding'' allows surface detectors to search for featureless signal events and provide meaningful signal statistics for various phenomenological studies.
We stress that the underlying idea is very generic and thus can be straightforwardly applied to any existing and future surface-based experiments, hence strongly encourage experimental collaborations to consider the methodology elaborated in this paper for their future data analyses.

\section*{Acknowledgments}
We thank Joshua Berger, Yanou Cui, Sowjanya Gollapinni, R. Craig Group, Roni Harnik, Athanasios Hatzikoutelis, Jia Liu, and Anne Schukraft for useful and insightful discussions.
SS appreciates the hospitality of Fermi National Accelerator Laboratory.
DK is supported by the Korean Research Foundation (KRF) through the CERN-Korea Fellowship program.
KK is supported in part by US Department of Energy under grant no. DE-SC0017965.
JCP is supported by the National Research Foundation of Korea (NRF-2016R1C1B2015225) and the POSCO Science Fellowship of POSCO TJ Park Foundation.
SS is supported by the National Research Foundation of Korea (NRF-2017R1D1A1B03032076).

\bibliographystyle{JHEP}
\bibliography{Earth-Shield.bib}

\end{document}